\documentclass[journal,twoside,web]{ieeecolor}

\usepackage{tmi}
\usepackage{cite}
\usepackage{amsmath,amssymb,amsfonts}
\usepackage{algorithmic}
\usepackage{graphicx}
\usepackage{textcomp}
\usepackage{url}
\usepackage{multirow}%
\usepackage{manyfoot}%
\usepackage{mathrsfs}%
\usepackage{mathtools}%
\usepackage{booktabs}%
\usepackage{subfig}
\usepackage{xcolor}
\usepackage[colorlinks=true]{hyperref}

\newcommand{\RR}[0]{\mathbb{R}} 
\newcommand{\ve}[1]{\mathbf{#1}} 
\newcommand{\grads}{{\nabla^\text{S}}} 

\def\BibTeX{{\rm B\kern-.05em{\sc i\kern-.025em b}\kern-.08em
    T\kern-.1667em\lower.7ex\hbox{E}\kern-.125emX}}
\begin{document}
\title{SynthAorta: A 3D Mesh Dataset of Parametrized Physiological Healthy Aortas}
\author{Domagoj Bo\v{s}njak, Gian Marco Melito, Richard Schussnig, Katrin Ellermann, and Thomas-Peter Fries
\thanks{This paragraph of the first footnote will contain the date on which
you submitted your paper for review. This work was in part funded by Graz University of Technology, Austria, through the LEAD Project on ``Mechanics, Modeling, and~Simulation of Aortic Dissection''. R.S. was partially supported by the German research foundation through project no. 524455704 (``High-Performance Simulation Tools for Hemodynamics'') and partially supported by the EuroHPC joint undertaking Centre of Excellence dealii-X, grant agreement no. 101172493.}
\thanks{Domagoj Bo\v{s}njak and Thomas-Peter Fries are with the Institute of Structural Analysis, Graz University of Technology, Lessingstrasse 25, Graz 8010, Austria (e-mail: bosnjak@tugraz.at; fries@tugraz.at).}
\thanks{Gian Marco Melito and Katrin Ellermann are with the Institute of Mechanics, Graz University of Technology, Kopernikusgasse 24/IV, Graz 8010, Austria (e-mail: gmelito@tugraz.at; ellermann@tugraz.at).}
\thanks{Richard Schussnig is with the Faculty of Mathematics, Ruhr University Bochum, Universit\"{a}tsstrasse 150, Bochum 44780, Germany (e-mail:richard.schussnig@rub.de)}
}

\maketitle

\begin{abstract} 
The effects of the aortic geometry on its mechanics and blood flow, and subsequently on aortic pathologies, remain largely unexplored. The main obstacle lies in obtaining patient-specific aorta models, an extremely difficult procedure in terms of ethics and availability, segmentation, mesh generation, and all of the accompanying processes. Contrastingly, idealized models are easy to build but do not faithfully represent patient-specific variability. Additionally, a unified aortic parametrization in clinic and engineering has not yet been achieved. To bridge this gap, we introduce a new set of statistical parameters to generate synthetic models of the aorta. The parameters possess geometric significance and fall within physiological ranges, effectively bridging the disciplines of clinical medicine and engineering. Smoothly blended realistic representations are recovered with convolution surfaces. These enable high-quality visualization and biological appearance, whereas the structured mesh generation paves the way for numerical simulations. The only requirement of the approach is one patient-specific aorta model and the statistical data for parameter values obtained from the literature. The output of this work is \emph{SynthAorta}, a dataset of ready-to-use synthetic, physiological aorta models, each containing a centerline, surface representation, and a structured hexahedral finite element mesh. The meshes are structured and fully consistent between different cases, making them imminently suitable for reduced order modeling and machine learning approaches.
\end{abstract}


\section{Introduction}
\label{sec:introduction}

\IEEEPARstart{C}{omputational} methods for the modeling of the vascular system have been growing in popularity in recent years, as they offer a plethora of possibilities for clinicians and engineers alike. A non-exhaustive list includes modeling of digital twins or virtual cohorts, supporting physicians in making assessments and predictions, as well as analyzing the effects of various parameters on target mechanical quantities, and consequentially on the health of the patients. An aspect of particular interest, which indisputably has a crucial impact on the mechanics of the aorta, is its \emph{geometry}. {This is intuitively clear, and verified by multiple studies, such as stroke propensity analysis by Choi~et~al.~\cite{Choi_2017a} and the geometry-based analysis of new-onset heart failure by Beeche~et~al.~\cite{Beeche_2024a}. Morphological studies have demonstrated that Type B aortic dissection patients exhibit significantly increased aortic arch dimensions, with longer arch lengths~\cite{Qiu2020}, whereas type A aortic dissection can potentially be predicted from abnormalities in the ascending aorta geometry~\cite{DellaCorte2021}. Minderhoud~et~al.\cite{Minderhoud2024} argued that patients with a repaired aortic coarctation exhibit a smaller aortic arch, suggesting geometric features could be of importance in long-term risk assessment. As the outlined works verify, geometric parameters play a key role in understanding mechanics and patient health, though their effects remain underexplored.}

Many works have {studied the aortic geometry and its effects}~\cite{PrahlWittberg_2015a,Marrocco-Trischitta2017,Bruse2017,Sievers2017,Gounley_2017a,Saitta2022a}. However, a barrier that is ever-present lies in obtaining usable patient-specific models. Existing works usually make use of a small number of patient-specific geometries. This is completely understandable when one considers the difficulties and the user interaction required for obtaining computational models from patient-specific CT scans~\cite{Zhang_2007a,Ghaffari_2017a,Bosnjak_2023a,Decroocq_2023a}. Besides, obtaining the CT scans themselves already presents a challenge~\cite{Antiga_2002b, Zhong_2018a}.

\begin{figure*}[!ht]
    \centering
    \includegraphics[width=\linewidth]{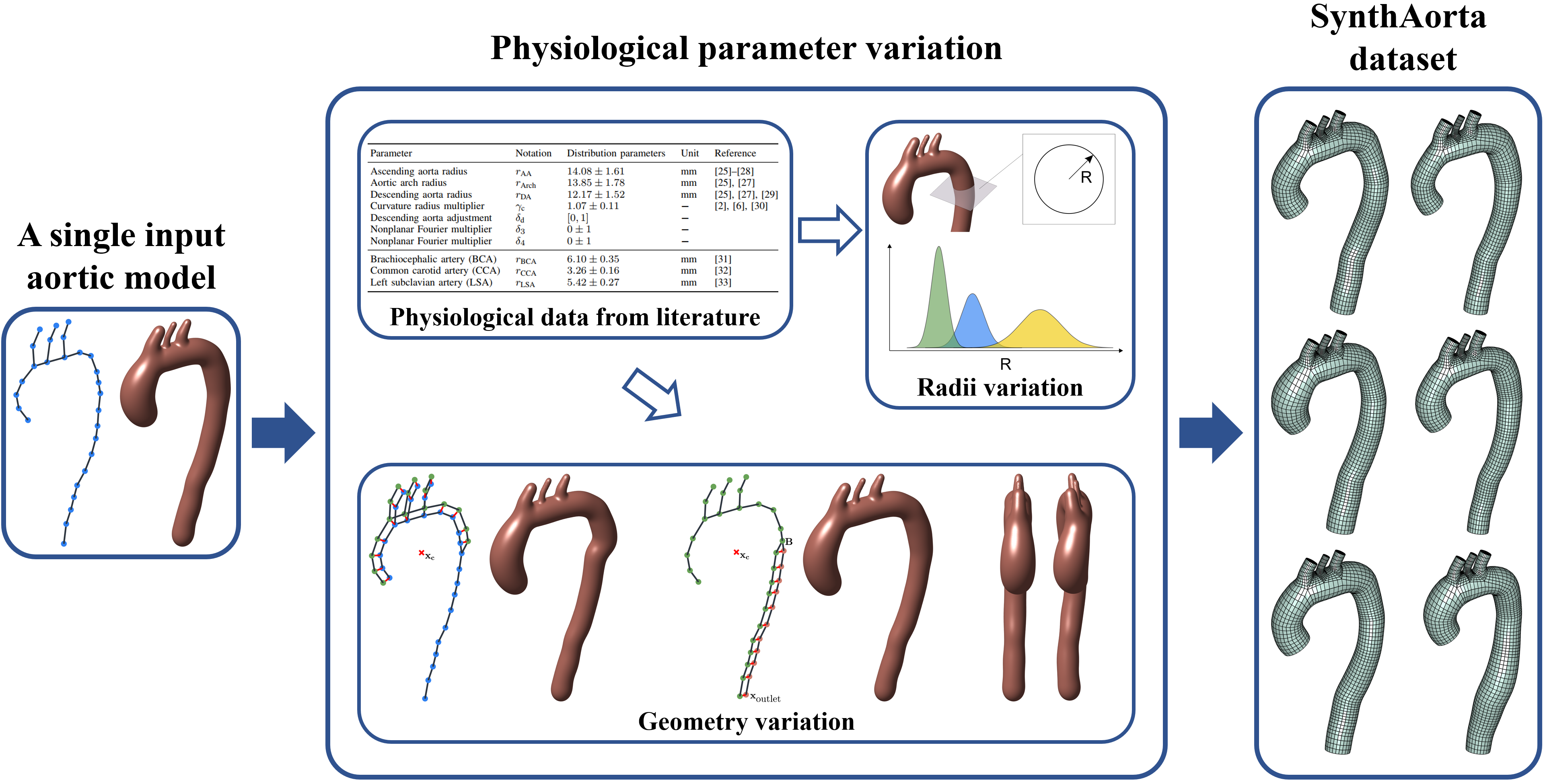}
    \caption{The overview of the SynthAorta approach: starting from a single base model, physiologically motivated geometrical and radius parameters are modeled statistically, and varied on the level of the coarse skeleton. The surface is reconstructed purely from the skeleton using convolution surfaces, enabling the generation of a dataset of structured hexahedral meshes, which can be used in numerical simulations.}
    \label{fig:enter-label}
\end{figure*}

As an alternative, several approaches are based on (parametrized) idealized models of the aorta for the exploration of geometrical parameters~\cite{Vasava_2012a,BenAhmed_2016a, Cilla_2020a, Jafarinia2023}. However, idealized models might fail to faithfully represent patient-specific behavior due to their inherent simplicity. Moreover, the parameters often do not correspond to clinically measurable quantities. Finally, idealized models often lead to meshes with sharp corners, which impact numerical simulations in a non-physiological manner. 

These obstacles have been the main motivators of this work, in which we seek to generate parametrized, synthetic, but also smooth and physiological models of the human aorta. A systematic analysis of the effects of the aortic geometry could be made much more streamlined and efficient if one had a database of parametrized models at hand. Nonetheless, even if we turn to virtual models, a significant barrier is a lack of a systematic parametrization of the aortic geometry~\cite{Khabaz_2024a}. 

Proper statistical and uncertainty modeling combined with physiological soundness enables easy analyses of various effects on arbitrary mechanical quantities of interest. The introduced parametrization is suitable for a wide array of engineering studies, including blood flow simulations~\cite{Melito2021,Ranftl2021,Pacheco2021a}, investigations into thrombus formation~\cite{Melito2020,Jafarinia2023,Badeli2023,Schussnig2022b}, and analyses of fluid-structure interaction~\cite{Schussnig2021a,Schussnig2022c,Baeumler2020,Schussnig2021d, Schussnig_2024a}. 
In this work, we define a new set of parameters to mathematically and statistically model the geometry of the aorta. The main features include a clear physiological meaning, measurability, mutual independence, and the fact that their stochastically sampled values are sourced from patient-specific data whenever possible~\cite{Redheuil2011,Barker2012,Chang2020,Wu2023,Wolak2008,Marrocco-Trischitta2017,Boufi2017,Saitta2022a,Panagouli2020,Barral2011,Schaefer2024a}. 
Its implementation is designed to be versatile, closely reflecting clinical quantities as they are measured in clinical settings. The proposed approach enhances the usability and comprehension of the geometry, avoiding the use of complex and unphysiological parameters that may arise from methods such as principal component analysis~\cite{Cosentino2020,Schafer2023,Wiputra2022}. Although these methods can effectively isolate and describe intricate geometrical features, they often need more transferability to the clinical setting due to the absence of physiological and physically comparable measures.

Imposing a parametrized model to generate (smooth) aortic geometries presents a challenging task. In our approach, we only require \emph{a single base model}, containing a centerline which consists only of points and straight segments. Then, the aforementioned parameters are applied only to the centerline, which is an extremely simple object. While it is clear that a centerline may be extracted from the surface of a patient-specific aorta, going the other way around may not seem like a viable task. Here, we make use of \emph{convolution surfaces}~\cite{Bloomenthal_1997a,FuentesSuarez_2019a}, a graphics methodology to render surfaces based purely on centerline and radial data. Essentially, a small number of points, segments, and associated radii (order of magnitude of \(20-30\)) are enough to describe a smooth surface of an aorta. Thus, parametrizing the centerline and radius information enables the parametrization of the entire aorta. {The base model and centerline are obtained from~\cite{Bosnjak_2023a}, where the centerline was computed based on the mesh contraction approach~\cite{au2008skeleton}.}

Even though the generated surfaces are useful, they are still insufficient to perform numerical simulations. The primary goal herein is to enable users to 
directly employ the finite element, finite volume, finite difference or any other mesh-based numerical method on high-quality meshes to effectively analyze the physical processes at work.
With this objective in mind, it is necessary to produce volumetric discretizations of the generated aortic surfaces, i.e., \emph{meshes}. The standard methodology for non-trivial domains such as aortas is \emph{unstructured meshing}~\cite{Owen_2000a}. This is justified by the flexibility and robustness of the meshing procedure, as well as a multitude of off-the-shelf software tools~\cite{Geuzaine_2009a,Si_2015a}. On the other hand, it is often worth investing extra effort into the generation of structured meshes~\cite{Armstrong_2015a, Bosnjak_2023a}, as they offer certain advantages. Examples include exact control of the element count, straightforward formation of boundary layers, flow-oriented elements, and finally, the control of the local mesh structure. Hence, the resulting meshes have the same node placement independently of the geometry variation, greatly boosting their applicability for various applications such as machine learning or reduced order modeling. To realize the mesh generation, we make use of the block-structured approach presented in~\cite{Bosnjak_2023a}, extended in \cite{Bosnjak_2023b, Bosnjak_2024a}. It fits well with the approach described herein since the required input is exactly a convolution surface description of the domain. 

{Ultimately, this dataset presents the main step towards enabling geometry-based analyses of the aorta. The main barrier towards devising a machine learning or reduced order model that can reach conclusions on mechanical quantities purely from geometrical information, without performing new simulations, is in fact the lack of available simulation data. Performing numerical simulations on the high-quality meshes in this dataset provides exactly the key data necessary for such models.}

\section{Methods}
In this section, we present the surface representation methodology, the parametrization model of the aorta, and briefly the mesh generation procedure. The base patient-specific model, now referred to as the \emph{base model}, is employed here as a starting point for the generation of the parametrized aorta. The goal is to produce geometries that do not heavily depend on or resemble the base model; it is taken as patient-specific simply to have a reasonable starting point. The produced parameters are sampled using a Latin hypercube sampling method, ensuring a comprehensive parameter space exploration. The generation of the aortas in the dataset follows this procedure, with the sampled points adhering to physiological statistical distributions collected from the available literature.

\subsection{Convolution surfaces}
The surface of the domain is represented implicitly using the convolution surface approach of Fuentes Suárez~\cite{FuentesSuarez_2019a}. Benefits of the approach include surface smoothness, and natural blending between vessels. The starting point is a centerline which consists of a set of points and segments connecting them, where each point is assigned a radius. We first consider a single centerline segment, defined by points \(\ve{a}, \,\ve{b} \in \RR^3\). The associated convolution surface function \(C_{\Gamma}^{K}: \mathbb{R}^3 \rightarrow \mathbb{R}\) is defined as
\begin{equation}
    C_{\Gamma}^{K}(\mathbf{x}) = \int_0^{1} K \left( g\left(  \Gamma(s), \mathbf{x} - \Gamma(s) \right) \right) g\left(\Gamma(s), \Gamma'(s)\right) \mathrm{d}s,
\end{equation}
where \(\Gamma: [0,l] \rightarrow \RR^3\) denotes the parametrization of a centerline segment \([\ve{a},\, \ve{b}]\) {of length \(l\)} defined as
\(
    \Gamma_S(s) = \ve{a} + \frac{s}{l}(\ve{b}-\ve{a}).
\)
The compactly supported kernel function \(K: \RR \rightarrow \RR\) is defined as
\begin{equation}
    K(x) = 
    \begin{cases}
    \frac{35}{16} \left(1-x^2\right)^3, & \quad x \in [0,1]\\ 
    0,                      & \quad \text{otherwise}
    \end{cases}
    ,
\end{equation}
and the distance function \(g : \RR^3 \times \RR^3 \rightarrow \RR\) as
\(
    g(\ve{x}, \ve{y}) = \sqrt{\ve{x}^T\,G(\ve{y})\,\ve{x}}.
\)
{The matrix \(G\) is computed in an eigenvalue decomposition form
\(
    G = UDU^T.
\)}
Following~\cite{FuentesSuarez_2019a}, \(U\) is obtained by taking the rotation matrix that maps the vector \(e_1=(1,0,0)\) to the tangent vector of the respective segment \(\Gamma'(s).\) The matrix \(D\) captures the radius information using \(D(s) = \textrm{diag}(\alpha(s), \beta(s), \gamma(s))\), controlling the linearity of the radius along the segment. The functions \(\alpha,\beta \textrm{ and } \gamma\) are computed from the equation \(\xi(s) = \big( \frac{l-s}{l}\xi_0^{-\frac{1}{2}} +  \frac{s}{l} \xi_1{-\frac{1}{2}} \big)^{-2}\), where the distinct initial values \(\xi_0\) and \(\xi_1\) for each of the three functions are computed from the radiuses of the segment \(r_{\ve{a}}\) and \(r_{\ve{b}}\). This framework enables anisotropic convolution surfaces with non-circular cross-sections. We omit further details for brevity, as this construction mainly serves to prove the radius control as shown in~\cite{FuentesSuarez_2019a}.

To compute the full surface representation function, we sum over all of the segments of the centerline denoted by \(\mathbb{S}\):
\begin{equation}
    C^K(\ve{x}) = \sum_{S\in\mathbb{S}} C_{\Gamma_S}^K(\ve{x}).
\end{equation}

We refer the reader to \cite{FuentesSuarez_2019a} for more details. The main advantages of this concrete convolution surface choice are the provably good radius control and the fact that the kernel function \(K\) is compactly supported, meaning that the evaluation of the total convolution surface function can be greatly sped up. 

\subsection{Statistical geometric parametrization} 
Two groups of geometrical parameters generate a set of virtual healthy aortas. Group A identifies the parameters that are actively tuned for the creation of new aortas. In contrast, group B varies due to the changes imposed on the first group. The geometrical parameters are shown in Table \ref{tab:aorticParametersDistributions}. We performed an extensive literature review on the geometrical characterization of measurable aortic parameters. This review allowed us to define each geometrical parameter's probability distribution functions to create physiological results. 

The development of a statistical geometric parametrization of a healthy aorta was driven by the need to define measurable clinical quantities. However, specific parameters had to be imposed to account for a wide range of shapes for modeling purposes.
Based on a comprehensive literature review, our findings indicate that specific geometrical parameters, such as the descending aorta slope ($\rho$) and the two nonplanar Fourier series multipliers ($\delta_3$ and $\delta_4$) defined in Sect.~\ref{sec:descendAortAdj} and Sect.~\ref{sec:fourierDispl} respectively, do not correspond to a clinically measurable quantity. Nonetheless, they represent essential geometrical aortic features. Therefore, we have mathematically modeled these parameters to illustrate important geometrical features of the aorta.

\begin{table*}[htbp]
  \centering
  \caption{Aortic parameters and their probabilistic distributions. The parameters in group A are purposefully modified to create new aortic datasets. The parameters in group B are produced as a result of the newly generated aorta. The distribution parameters, e.g., $(a,b)$, are defined as mean and variance for a normal distribution, as location and scale for a Gumbel distribution, and as minimum and maximum values for a uniform distribution. Some of the parameter combinations produce invalid or unphysiological domains, causing the convolution surface to blend, in particular between the carotid arteries. In a minority of the cases, the meshing algorithm produces invalid elements. Thus, the final distributions are slightly affected. {The units in the SynthAorta statistical distributions are preserved as in the literature.}}
  \begin{tabular}{lllccllccc}
    \toprule
     & & & \multicolumn{2}{c}{Collected distribution} & & & \multicolumn{2}{c}{SynthAorta distribution} \\
    \cmidrule(lr){4-5}\cmidrule(lr){8-9}
    Group & Parameter & Notation & \multicolumn{1}{c}{type} & \multicolumn{1}{c}{parameters} & Unit & Reference & \multicolumn{1}{c}{type} & \multicolumn{1}{c}{parameters} \\
    \midrule
    \multirow{7}[2]{*}{A} & Ascending aorta radius & $r_\text{AA}$ & Normal     & $(14.08, 2.59)$ & mm         & \cite{Redheuil2011,Barker2012,Chang2020,Wu2023} & Normal     & $(13.69, 2.30)$ \\
    
   & Aortic arch radius & $r_\text{Arch}$ & Normal     & $(13.85, 3.16)$ & mm & \cite{Redheuil2011,Chang2020} & Normal     & $(13.00, 1.99)$ \\
   
   & Descending aorta radius & $r_\text{DA}$ & Normal     & $(12.17, 2.31)$ & mm & \cite{Wolak2008,Redheuil2011,Chang2020} & Normal     & $(12.17, 2.31)$ \\
   
   & Curvature radius multiplier & $\gamma_\text{c}$ & Normal     & $(1.07, 0.01)$ & $-$ & \cite{Marrocco-Trischitta2017,Boufi2017,Saitta2022a} & Gumbel     & $(1.10, 0.06)$ \\
   
   & Descending aorta adjustment & $\rho$ & & & $-$ & & Uniform & $(0, 1)$ \\
   
   & Nonplanar Fourier multiplier & $\delta_3$ & & & $-$ & & Normal & $(1, 0.09)$ \\
   
   & Nonplanar Fourier multiplier & $\delta_4$ & & & $-$ & & Normal & $(1, 0.09)$ \\
   
    \midrule
    \multirow{3}[2]{*}{B} & Brachiocephalic artery (BCA) radius & $r_\mathrm{BCA}$ & Normal     & $(6.10, 0.12)$ & mm         & \cite{Panagouli2020} & Normal     & $(6.10, 0.12)$ \\
    
   & Common carotid artery (CCA) radius & $r_\mathrm{CCA}$ & Normal     & $(3.26, 0.03)$ & mm         & \cite{Schaefer2024a} & Normal     & $(3.26, 0.03)$ \\
   
   & Left subclavian artery (LSA) radius & $r_\mathrm{LSA}$ & Normal     & $(5.42, 0.07)$ & mm         & \cite{Barral2011} & Normal     & $(5.42, 0.07)$ \\
   
   & Centerline curvature radius & $R_\text{c}$ & Normal     & $(39.30, 3.92)$ & mm         & \cite{Marrocco-Trischitta2017,Boufi2017,Saitta2022a} & Gumbel     & $(40.41, 2.40)$ \\
   
    \bottomrule
    \end{tabular}
  \label{tab:aorticParametersDistributions}%
\end{table*}%

\subsubsection{Aortic radii} \label{sec:aorticRadii}

\begin{figure}[h]
    \centering
    \subfloat[]{
        \includegraphics[width=0.42\linewidth]{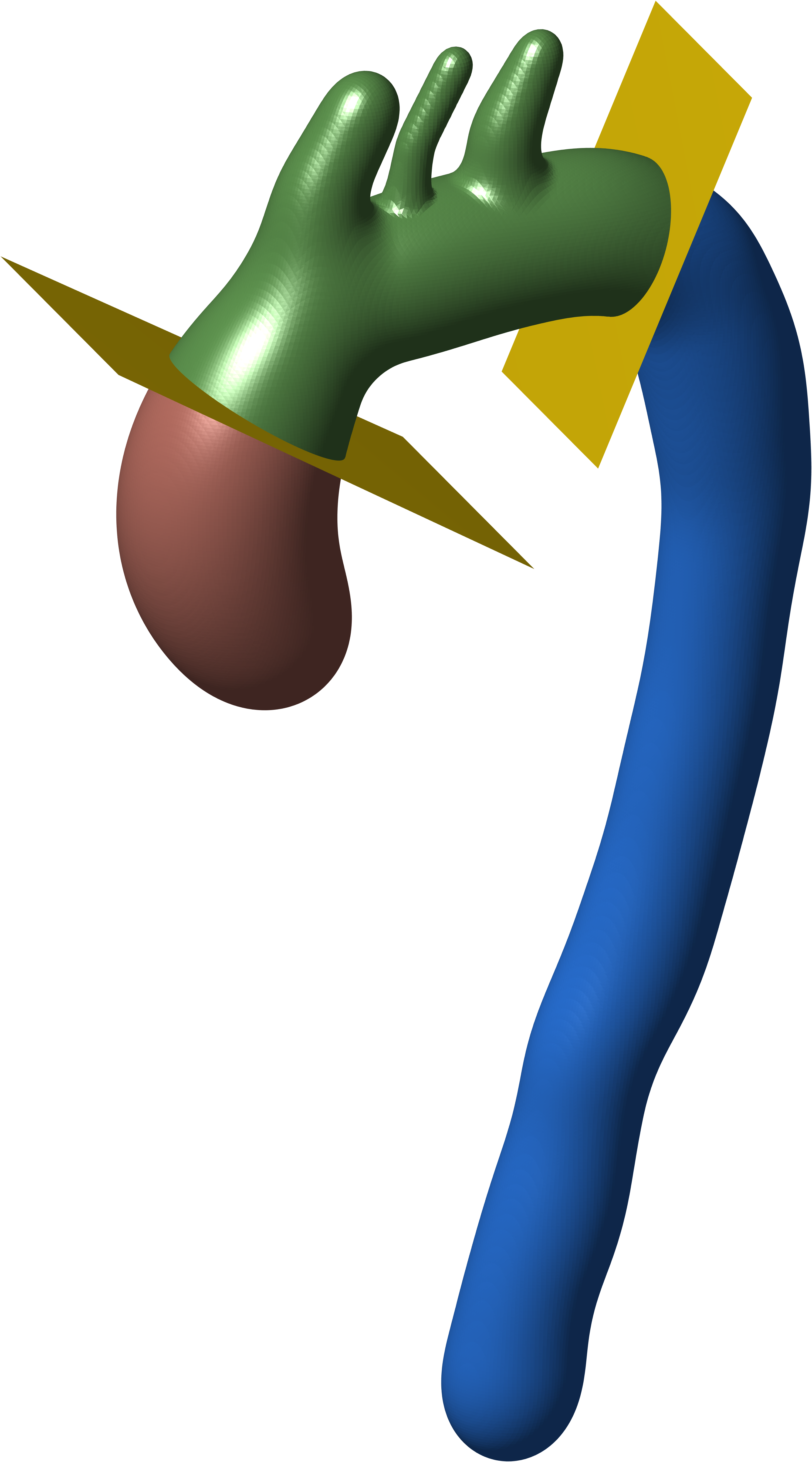}
    }
    \hspace{1.0cm}%
    \subfloat[]{
        \includegraphics[width=0.3\linewidth]{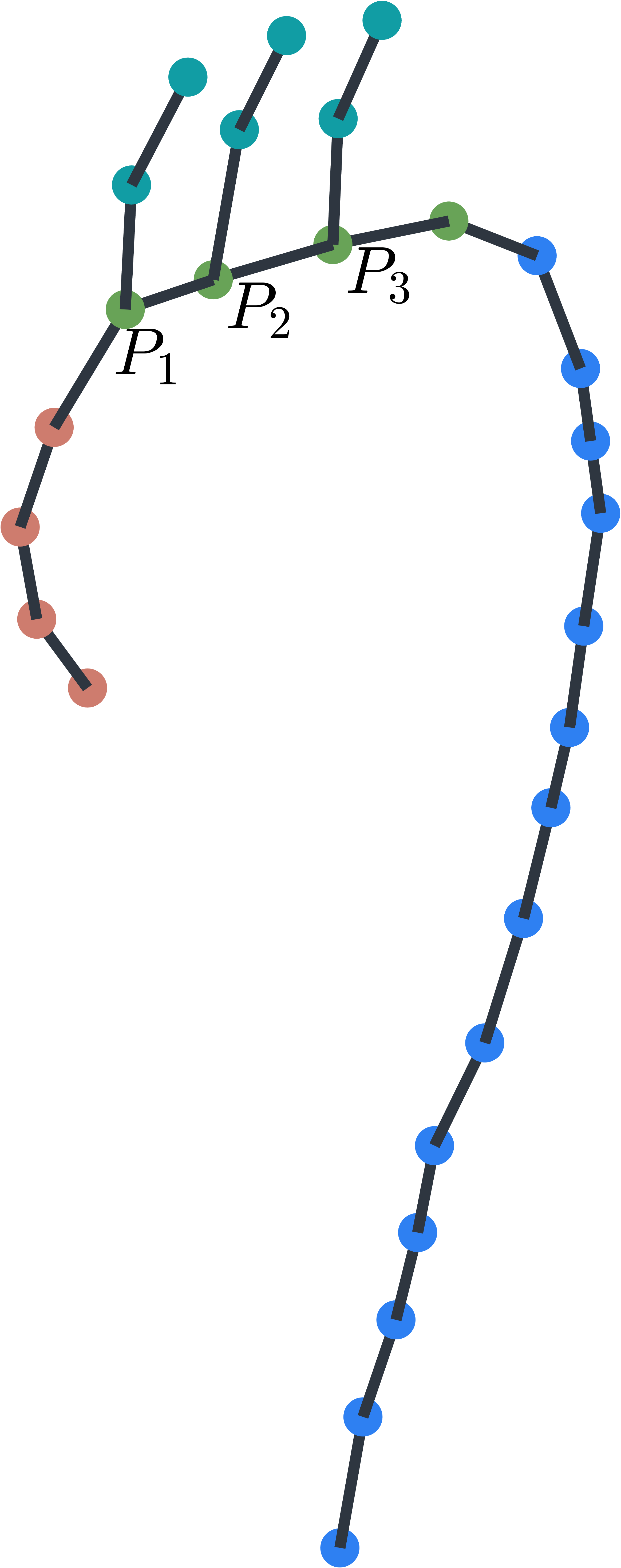}
        \label{fig:aorticRadii_CommonPts}
    }
    \caption{(a) Schematic representation of the aorta, divided into three segments: the ascending aorta, the aortic arch, and the descending aorta. (b) The aortic centerline divided into the three aortic segments. The points $P_1$, $P_2$, and $P_3$ mark the locations where the carotid arteries intersect with the aortic path.}
    \label{fig:aorticRadii}
\end{figure}

The aorta is often divided into three segments for clinical analysis: the ascending aorta, the aortic arch, and the descending aorta. This division aligns with extensive clinical literature documenting variations in aortic radii, aiding in the definition of normal and pathological conditions. By examining a broad set of scientific articles~\cite{Wolak2008,Redheuil2011,Barker2012,Chang2020,Wu2023}, we identified ranges of aortic radii that accurately represent a healthy aorta in all segments. Particular emphasis was placed on statistically distributed values wherever available, as the use of a probabilistic distribution function enhances the parametrization of the aorta. This approach creates a statistical domain of healthy aortas, enabling the generation of new aorta samples within this domain. An example of the subdivision of the aorta is shown in Fig.~\ref{fig:aorticRadii}.

Our literature review, see Table~\ref{tab:aorticParametersDistributions}, revealed that some studies provide probabilistic distributions of aortic diameters or radii, typically defined by mean values and standard deviations. These are often implicitly considered as normally distributed variables. Other studies report aortic diameters or radii as a range of values, suggesting uniformly distributed measurements. However, it is important to note that the collected values in the literature almost never report the exact statistical distribution of the collected data. They are always summarized with mean and standard deviation or minimum and maximum values. This highlights a significant lack of detailed distribution definitions, which can often differ, e.g., be multimodal. Our study incorporates normally distributed radii for each aortic segment to close this gap, ensuring a comprehensive mathematical definition of aortic geometrical variations.

The collected data is also strongly heterogeneous. Some studies do not divide the aorta into ascending, arch, and descending aorta, as used in our study. Others classify geometries by age or sex. For our purposes, we combined all measurements into a single database, disregarding age and sex distinctions. This decision was made to create a database representing the full spectrum of healthy aortic morphologies. 

To account for different population sizes across the $S$ reviewed studies, we employed a weighted mean and a weighted standard deviation to compute the final mean and standard deviation of aortic radii in the three aortic segments. The weighted values are defined as
\begin{equation}
    \Theta_\text{w} = \frac{\sum_{\text{q}=1}^\text{S} \Theta_\text{q} w_\text{q}}{\sum_{\text{q}=1}^\text{S} w_\text{q}},
\end{equation}
where $\Theta$ represents either the mean $\mu_\text{q}$ or the standard deviation $\sigma_\text{q}$ from the q-th study. The weights $w_\text{q}$ for each study are calculated by dividing the number of measured individuals by the total population of each aortic segment. We employ the computed weighted mean and standard deviation to generate new aortic geometries from the aortic radii of the base model. 

We use a Gaussian random field to create realistic radii samples of the three aortic segments, using a Karhunen-Loève (KL) expansion. {It is used only for computational efficiency, since many realizations of the random field are needed. A typical SVD truncation happens in the background, yielding acceptable accuracy while greatly improving efficiency.} This method helps us generate smooth, interconnected variations along the length of the aorta~\cite{Constantine2024}. 

The correlation structure is manually adjusted to reflect the physiological connections between the different segments of the aorta, ensuring that the produced aortas maintain realistic anatomical characteristics. We determine the average and variability of the random field from weighted values obtained from medical literature. These variations introduce physiological differences within each realization. The mean and variance of the random field are defined as $\mu_\text{w}$ and $\sigma_\text{w}$ and are derived from weighted values reported in the medical literature, see Table~\ref{tab:aorticParametersDistributions}.

\subsubsection{Carotid arteries radii}

Following the data obtained from Sch\"afer et al.~\cite{Schaefer2024a} and the works in~\cite{Panagouli2020,Barral2011} we adjusted the carotid arteries' radii following the collected statistical distributions. The resulting carotid radii are applied to each entire carotid artery radii.

\subsubsection{Centerline curvature radius}
\begin{figure}[t]
    \centering
    \subfloat[]{
        \includegraphics[width=0.30\linewidth]{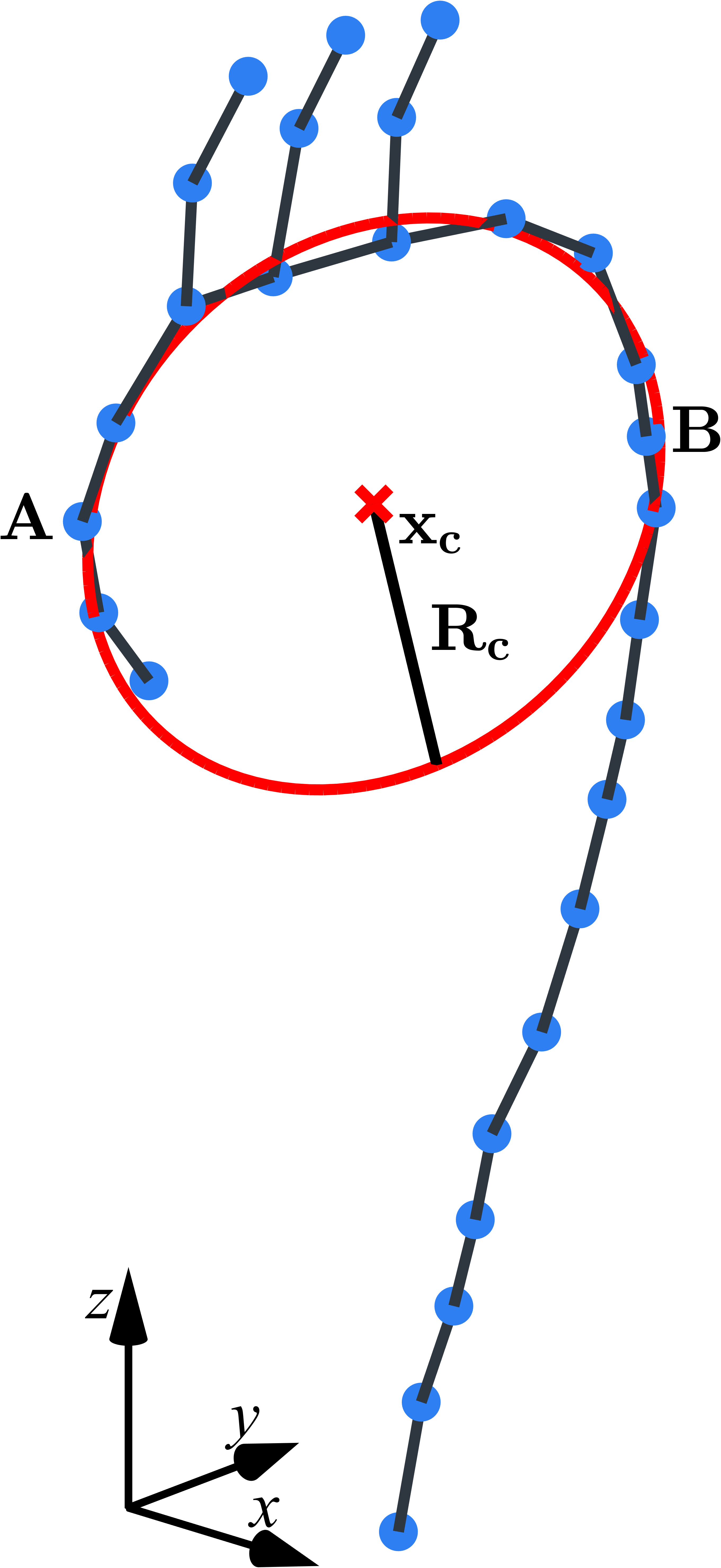}
        \label{fig:FlatcenterlineBase}
    }
    \hspace{0.0cm}%
    \subfloat[]{
        \includegraphics[width=0.30\linewidth]{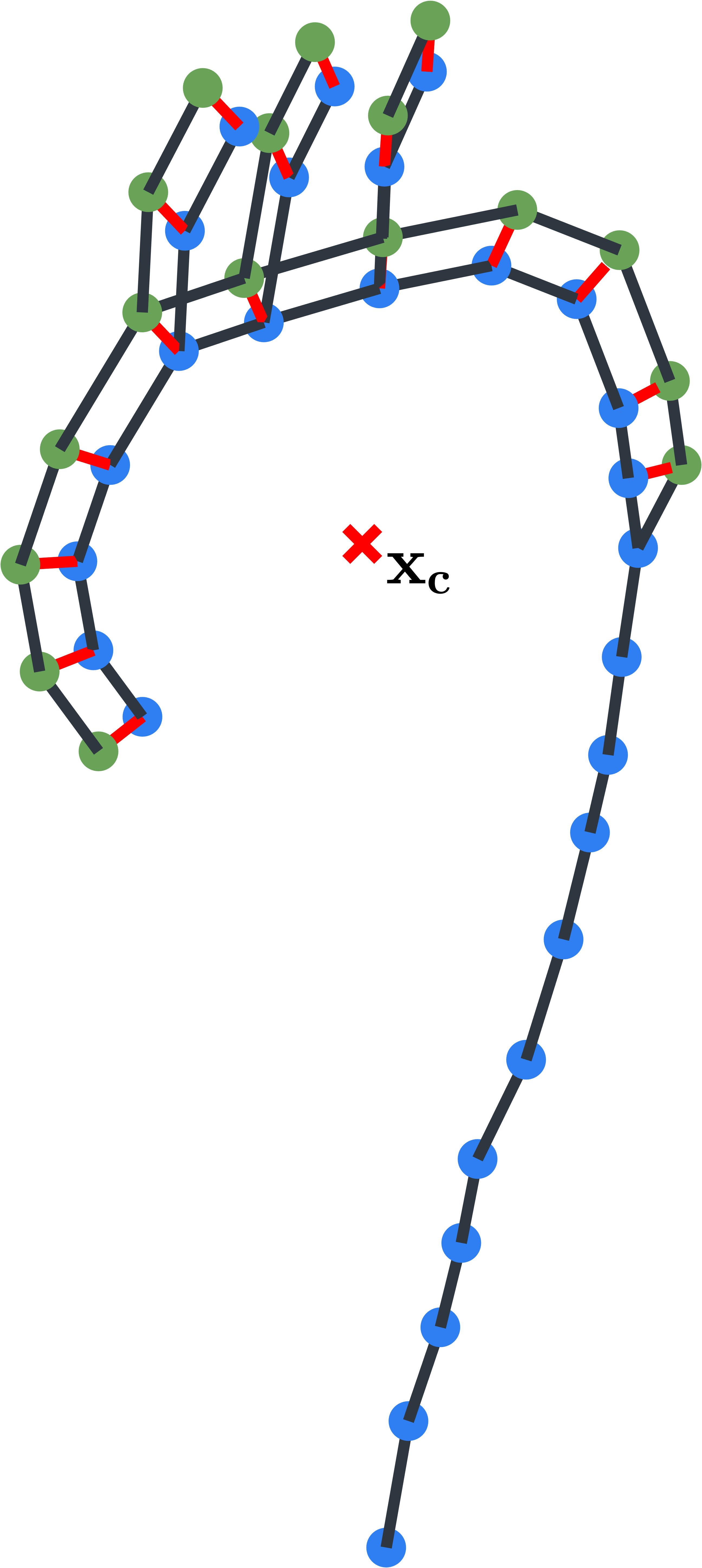}
        \label{fig:FlatcenterlineRcChange}
    }
    \hspace{0.0cm}%
    \subfloat[]{
        \includegraphics[width=0.31\linewidth]{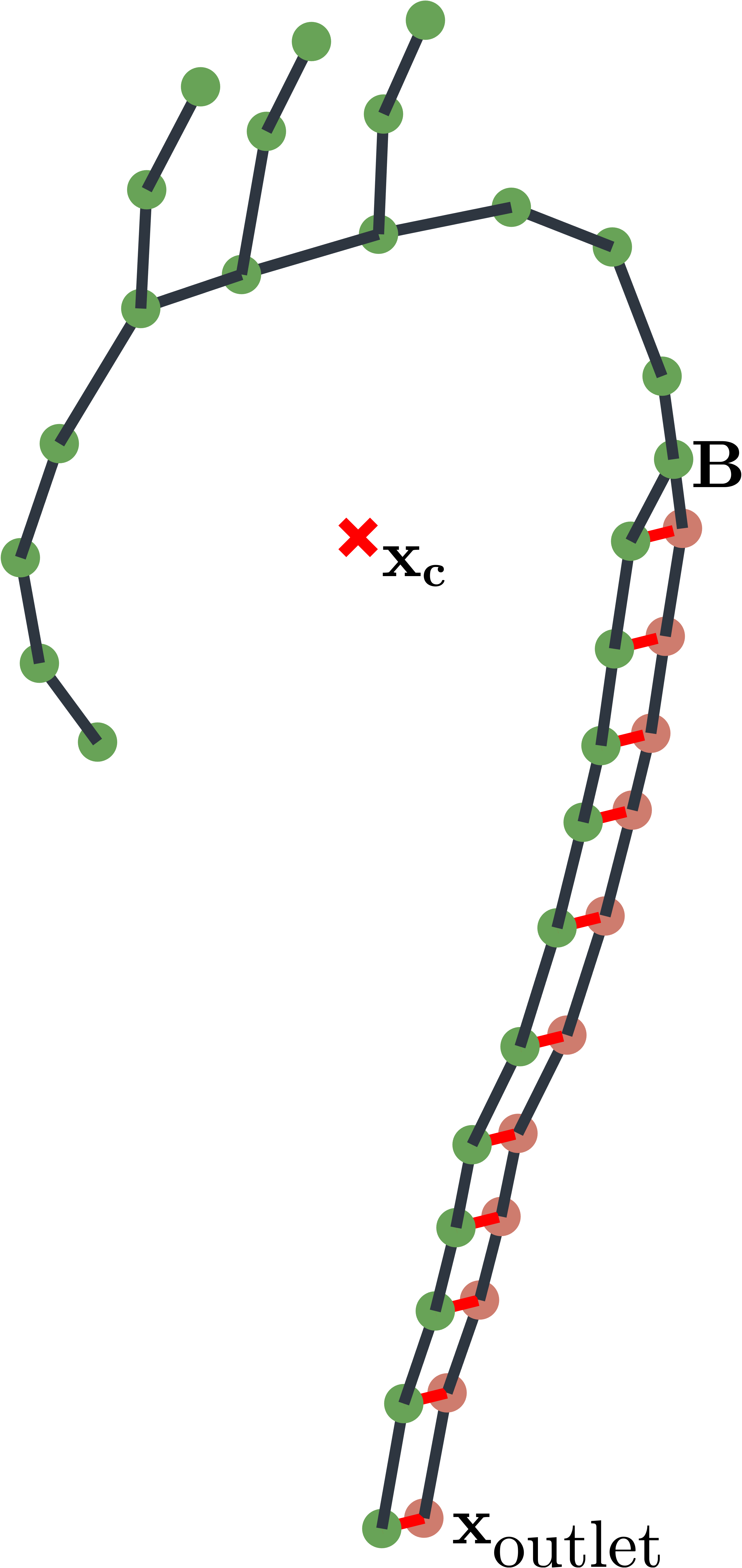}
        \label{fig:FlatcenterlineDescChange}
    }
    \caption{Centerline curvature radius and descending aorta adjustment parameters on the \emph{plane-projected} base centerline: (a) The locations of \(\ve{A}\) and \(\ve{B}\), as well as the visualization of the centerline curvature radius \(R_\text{c}\) and the center point \(\ve{x}_{\text{c}}\), (b) the variation in the centerline curvature radius modifies the points in the upper part of the aorta, in the direction of the center \(\ve{x}_{\text{c}}\), and (c) the correction of the previously unmodified points.}
    \label{fig:Flatcenterline}
\end{figure}

For the rest of the section, we assume the base centerline has been mapped to a single best-fitting plane using a typical SVD-based procedure. {This is done to keep the definition of the centerline curvature radius in line with existing work~\cite{Saitta2022a}.} We define the point \(\ve{A}\) in the ascending aorta, {heuristically, as the point approximately closest to the bifurcation of the nearby pulmonary vessel, and consequentially} \(\ve{B}\) as the point on the centerline which is on the same axial plane as \(\ve{A}\), the same as in \cite{Marrocco-Trischitta2017} and \cite{Saitta2022a}, shown in Fig.~\ref{fig:FlatcenterlineBase}.
Denoting the centerline points of the main aorta which lie between them: \(\ve{A} = \ve{x_1}, \ve{x_2}, \dots, \ve{x_{n-1}}, \ve{x_n} = \ve{B}\), the centerline curvature radius is determined as follows. For the initial patient-specific centerline, we determine the center point \(\ve{x_\text{c}} = (x_\text{c}, y_\text{c}, z_{\text{c}}),\) and the radius \(r\) of the circle by solving the following optimization problem
\begin{equation}
\underset{\ve{x_\text{c}}, r}{\min}\,\,{\sum_{i=1}^{n} \left( \left\| \ve{x_i} - \ve{x_\text{c}} \right\| - r \right)^2} ,   
\label{eq:nonlinearRadiusOptimization}
\end{equation}
using the Gauss-Newton optimization algorithm. Since it is not clear how to uniquely vary the centerline curvature radius under these conditions, we only perform this computation once. Then, the center of the circle \(\ve{x_\text{c}}\) is kept fixed, and only the radius is varied. In that case, it can be easily proven that the solution to the optimization problem that only depends on \(r\) is given by
\begin{equation} \label{eq:curvRadius}
r_{\text{best}} = \frac{1}{n}\sum_{i=1}^{n} \left\| \ve{x_i} - \ve{x_\text{c}} \right\|,    
\end{equation}
i.e., the sought-for radius is the arithmetic mean of the distances of the points to the fixed circle center. Thus, if we multiplicatively move the points in the direction of the center, by a coefficient \(\gamma_\text{c} > 0\):
\begin{equation}
    \ve{x_i} \leftarrow \gamma_{\text{c}}\ve{x_i} + \left(1 - \gamma_{\text{c}}\right)\ve{x_{\text{c}}},
    \label{eq:CentCurvRadPointMap}
\end{equation}
the sum in Eq.~\ref{eq:curvRadius} becomes
\begin{equation}
\frac{1}{n}\sum_{i=1}^{n} \left\| \gamma_\text{c}\left(\ve{x_i} - \ve{x_\text{c}}\right) \right\| = 
\gamma_\text{c} \left(\frac{1}{n}\sum_{i=1}^{n}\left\| \ve{x_i} - \ve{x_\text{c}}\right\|\right) = \gamma_\text{c} r_{\text{best}},    
\end{equation}
meaning that the starting value of \(r\) only has to be multiplied by the same coefficient \(\gamma_\text{c}\). Thus, the centerline curvature radius \(R_\text{c}\) can be defined as:
\begin{equation}
    R_\text{c}(\gamma_\text{c}) = \gamma_\text{c} \, r,
\end{equation}
where \(r\) is the solution of the nonlinear optimization problem shown in Eq.~\ref{eq:nonlinearRadiusOptimization}, computed only once, from the base centerline. For completeness, the rest of the points in the ascending aorta are mapped as in Eq.\ref{eq:CentCurvRadPointMap}.

At this point, to impose a variation on the centerline curvature radius, the coefficient $\gamma_\text{c}$ can be considered as a random variable with a normal probability distribution function. This is defined to produce physiological centerline curvature radii in accordance with the reviewed literature. For some parameter values or combinations, the convolution surface blending~\cite{Bloomenthal_1997a} causes the carotid arteries to merge. Thus, the coefficient distribution is modified to produce only valid aortic surfaces; see Table \ref{tab:aorticParametersDistributions}. An example of increasing the centerline curvature radius \(R_\text{c}\) is shown in Fig.~\ref{fig:FlatcenterlineRcChange}.

\subsubsection{Descending aorta adjustment} \label{sec:descendAortAdj}
The variation of the centerline curvature radius \(R_\text{c}\) impacts the position of all the centerline points except the ones between the point \(\ve{B}\) and the main outlet point of the aorta. A new parameter is introduced to account for the analogous displacement of those points.

Let us consider the centerline points \(\ve{x_n},\,\dots, \ve{x_{n+k}}\), starting from \(\ve{x_n}=\ve{B}\) and ending at the main outlet point of the centerline \(\ve{x_{n+k}}=\ve{x_{\text{outlet}}}\), with the connections between neighbouring points \(\ve{x_j}\) and \(\ve{x_{j+1}}\). Let \(\Delta \ve{B} \in \RR^3\) be the displacement of the point \(\ve{B}\) resulting from a change in the centerline curvature radius \(R_\text{c}\). We introduce a new parameter \(\rho\) uniformly distributed between $0$ and $1$, defining the displacement at the outlet point \(\ve{x_{\text{outlet}}}\) as 
\begin{equation}
\Delta \ve{x_{\text{outlet}}} = \rho\Delta \ve{B},
\end{equation}
meaning it can range from no displacement to having the same displacement as a point \(\ve{B}\). Then, for the rest of the points \(\ve{x_j}\) in the descending part of the aorta, we perform a simple interpolation, resulting in the following displacement
%
\begin{equation}
{
\Delta \ve{x_j} = \Big((1-\rho)\Big(1-\frac{\|\ve{B} - \ve{x_j}\|}{\|\ve{B} - \ve{x_{\text{outlet}}}\|}\Big) + \rho\Big) 
\Delta \ve{B},
}
\end{equation}
{where \(j=n+1,\dots,n+k-1\).} The displacement at \(\ve{x_j}\) is a therefore a distance-based interpolation of displacements at \(\ve{B}\) and \(\ve{x_{\text{outlet}}}\). An example is illustrated in Fig.~\ref{fig:FlatcenterlineDescChange}.

\subsubsection{Nonplanar displacement} \label{sec:fourierDispl}
The aortic centerline data exhibit a nonplanar displacement compared to its planar projection, denoted by the vector $\ve{d}$. To quantify it, we use a Fourier series expansion defined as
\begin{equation}
    \ve{d}_N(\ve{x})=A_0 + \sum_{n=1}^N \left( A_n \cos \left(n w {\|\ve{x}\|}\right) + B_n \sin \left(n w {\|\ve{x}\|}\right) \right),
\end{equation}
where $\ve{x}$ is the position vector of the centerline points, $N\in\mathbb{N}$ is defined as the series harmonics or truncation level, $A_n$ and $B_n$ and the series coefficients or harmonic amplitudes, and $w$ can be interpreted as the series frequency.

The goal is to determine not only the parameters $A_n$ and $B_n$ that can accurately describe the nonplanar displacement with a high level of approximation but also the truncation level $N$. Therefore, different truncation levels of the Fourier series were tested to fit the displacement data $\ve{d}$.

As the truncation level increases, the number of coefficients increases by $2$. Despite the improvements in the fit for $N>2$, the most suitable series, forcing a balance between the number of fitting parameters and fitting error improvement expressed by $\Delta R^2$, is a Fourier series with a truncation level of $N=2$. More details are shown in Table \ref{tab:fittingFourier}.

\begin{table}[]
  \centering
  \caption{Goodness of fit metrics for different truncation levels $N$ of the Fourier series model for the nonplanar aortic displacement, along with the series coefficients of $\ve{d}_2$. The percentage incremental $\Delta R^2$ is computed as the difference between the $R^2$ value for each truncation level $N$ and the $R^2$ value for $N=2$.}
    \begin{tabular}{llllll}
    \toprule
     $N$ & Fitting       & SSE$^{\text{a}}$ & RMSE$^{\text{b}}$ & $R^2$ & $\Delta R^2$ \\
        & parameters    & & & & \\
    \midrule
    2          & 4          & 11.69      & 0.88       & 0.88 & - \\
    3          & 6          & 10.26      & 0.88       & 0.89 & 1.1\% \\
    4          & 8          & 2.48       & 0.47       & 0.97 & 10.2\% \\
    \midrule  
    \multicolumn{6}{l}{Series coefficients of $\ve{d}_2$} \\
    \midrule
    $A_0$ & $A_1$ & $B_1$ & $A_2$ & $B_2$ & $w$ \\
    $-0.352$ & $-0.798$ & $-0.453$ & $1.517$ & $2.699$ & $0.027$ \\
    \midrule  
    \multicolumn{6}{p{200 pt}}{$^{\text{a}}$ Sum of the squared estimate of errors, or residual sum of squares} \\
    \multicolumn{6}{p{200 pt}}{$^{\text{b}}$ Root mean square error}
    \end{tabular}%
  \label{tab:fittingFourier}%
\end{table}%

In order to produce a set of randomly displaced centerlines, we introduce variability into the coefficients $A_n$ and $B_n$ for $n=1,2$. This is achieved by multiplying the series coefficients with random variables drawn from a normal distribution with a unitary mean and standard deviation of $0.3$ to avoid negative values. These random variables, denoted as $\gamma_j$ for $j=1,...,4$, play a crucial role in creating the desired variation in the centerlines. By adjusting the values of $\gamma_j$, we can generate an unlimited number of centerlines while maintaining their average alignment with the in-plane centerline. While we allow for this variability, certain series coefficients, such as $A_0$ and $w$, remain constant. This is because $A_0$ does not influence the nonplanar displacement of the aortic centerline, and $w$ represents a fundamental characteristic of the collected centerline that should remain unaltered.

The resulting Fourier series for a random centerline $k$ is
\begin{align}
    \hat{\ve{d}}_2^{(k)}(\ve{x}) = A_0 & + A_1 \gamma_1 \cos(w {\|\ve{x}\|}) + B_1 \gamma_2 \sin(w {\|\ve{x}\|}) \nonumber\\ 
            & + A_2 \delta_3 \cos(2w {\|\ve{x}\|}) + B_2 \delta_4 \sin(2w {\|\ve{x}\|}).
\end{align}

A global sensitivity analysis is performed to identify the most influential parameters, reducing the number of variables to modify, thus also reducing model complexity. The analyzed output is the root mean square error(RMSE) between the fitted data $\ve{d}_N$ and the generated random curves $\hat{\ve{d}}_2^{(k)}$. Using a polynomial chaos expansion with the LARS method and polynomial degree truncation $4$~\cite{Blatman2011}, it was found that only two variables, $\delta_3$ and $\delta_4$, significantly influence the discrepancies between the base model and randomly generated centerlines. Therefore, these parameters are identified as the most sensitive and further included as geometrical parameters for the generation of virtual healthy aortas.

\subsection{Mesh generation}
The mesh generation is performed in a structured manner. The input of the algorithm is a convolution surface, i.e., a centerline with radius information at each point. The output is a structured hexahedral mesh, with the same connectivity matrix for every case. This implies that the connectivity matrix only has to be stored once, and each domain only needs to keep track of the nodal positions. Moreover, it is very straightforward to include connectivity matrices for coarser meshes as well, meaning that multiple meshes can be stored for a single aorta, effectively at no cost.

The algorithm builds a block-structure~\cite{Ali_2016a}, a coarse domain subdivision, based on the input centerline. The block-structure generation is based on various centerline configurations occurring in the aorta. {Then, a sub-mesh is generated in each of the blocks using transfinite interpolation~\cite{Solin_2003a}. In order to ensure conformity and matching between the blocks, points are first generated on the edges of the block-structure. Then, surface nodes are mapped to the domain's surface using a gradient-based projection. Transfinite interpolation is then used to compute the face points using the already-computed edge points, subsequently applying the surface mapping again where necessary. Finally, the points in the block interiors are generated using the computed face and edge points via transfinite interpolation, completing the point generation procedure.}

{Once the submeshes inside the blocks are generated, the task of defining a global connectivity matrix simply means converting local node indices to global indices. Conformity is ensured during the procedure itself, making the connectivity matrix generation straightforward. More details on the procedure may be found in~\cite{Bosnjak_2023a,Bosnjak_2023b}.} 

Even though a given mesh can be further refined, the key question is whether this refinement increases the geometrical representation accuracy. Simply subdividing the elements into smaller ones does not yield a different geometrical representation. In the more common framework, where the domain is represented by a triangle surface mesh, a hard limit is imposed on the geometrical accuracy with respect to the original domain. However, since the convolution surface applied here is smooth in a mathematical sense~\cite{FuentesSuarez_2019a}, the spatial refinement can also arbitrarily increase the domain representation accuracy, provided that the surface points are projected with the convolution surface function post-refinement.

\subsection{Vascular flow simulation setup}

In order to demonstrate the application-readyness of the SynthAorta dataset, we employ the cases contained within the dataset for vascular flow simulation. As any other investigation starting from the dataset, we formulate a complete boundary value problem targeting the physics of interest. The setup considered herein is identical to~\cite{Bosnjak_2025b}, which considers the Navier--Stokes equations for incompressible flow of a generalized Newtonian fluid (see, e.g., \cite{Schussnig2021b, Pacheco2021c} for their numerical treatment),
\begin{align*}
    \frac{\partial}{\partial t}\ve{u} 
    + 
    \ve{u} \cdot \nabla \ve{u}
    - 
    2 \nabla \cdot \left(\nu \nabla^\mathrm{S}\ve{u} \right)
    + 
    \nabla p
    &= 
    \ve{0}
    & \text{in }\Omega \times (0,T]
    ,
    \\
    \nabla \cdot \ve{u} 
    &= 
    0
    & \text{in }\Omega \times [0,T]
    ,
\end{align*}
with $\grads (\cdot) \coloneqq \frac{1}{2}\left[\nabla (\cdot) + \nabla (\cdot)^\top\right]$, fluid velocity $\ve{u}$, pressure $p$ and variable kinematic viscosity $\nu$.
The domain $\Omega$ is discretized for each individual case as contained within the SynthAorta dataset. The viscosity is governed by the Carreau model,
\begin{gather*}
    \nu (\dot\gamma) = \eta_\infty + (\eta_0 - \eta_\infty) \left[ 1 + (\lambda \dot\gamma)^2 \right]^{\frac{n-1}{2}}
    ,
    \quad
    \\
    \text{with} 
    \quad
    \dot{\gamma} := \sqrt{2 \nabla^\mathrm{S} \mathbf{u} : \nabla^\mathrm{S} \mathbf{u}}
    ,
\end{gather*}
capturing the shear-thinning behavior of blood
with physiological fluid parameters taken from~\cite{Ranftl2021}.

The boundary conditions incorporate i) a physiological flow rate from~\cite{Baeumler2020}, which is adopted from a different patient, ii) no-slip boundary conditions on the stationary vessel wall, and iii) a fixed set of Windkessel parameters tuned as described in~\cite{Bosnjak_2025b}, simultaneously targeting the flow splits reported in~\cite{Baeumler2020}, and realistic absolute pressure values at the abdominal aorta of $120~\text{mmHg}$ in systole and $75~\text{mmHg}$ in diastole~\cite{Mills1970}. This setup delivers physiological, but non-patient-specific flows. Fixing the boundary conditions, we can investigate the effect of geometric variation.

In the context of the present work, we show that the meshes contained in the SynthAorta dataset are of sufficient accuracy and quality for an application in vascular flow, to yield physiological results when combined with suitable simulation tools and data. To this end, we i) perform a convergence study on the relevant quantities of interest in vascular flows, and ii) provide the flow rates and pressures on each of the outlets together with the spatial average of the wall shear stress (WSS) in the descending aorta.

The simulations are performed using \href{https://github.com/exadg/exadg}{\texttt{ExaDG}}~\cite{ExaDGgithub} (see \cite{Fehn2021b} and \cite{Arndt2020}). The software is based on \texttt{deal.II}~\cite{dealII95} and its matrix-free infrastructure~\cite{KronbichlerKormann2012, KronbichlerWall2018}, and supports $hp$-multigrid preconditioners based on~\cite{KronbichlerWall2018, Fehn2020}. Within this setup, the construction of nested mesh hierarchies as provided within the SynthAorta dataset is vital to obtain a suitable geometric coarsening sequence within the $hp$-multigrid preconditioner, enabling sufficient resolution and large sampling sets.

\section{Results}
The main output of this work is SynthAorta, a dataset of parametrized, physiological models of the healthy human aorta. Each model consists of:
\begin{itemize}
    \item Unique set of clinical parameters that define the geometry
    \item Centerline of the domain
    \item Smooth convolution surface representation
    \item Simulation-ready structured hexahedral mesh, with \(3\) additional coarser mesh levels 
\end{itemize}

\begin{figure*}[h!]
    \centering
    \subfloat[]{
        \includegraphics[height=3.5cm]{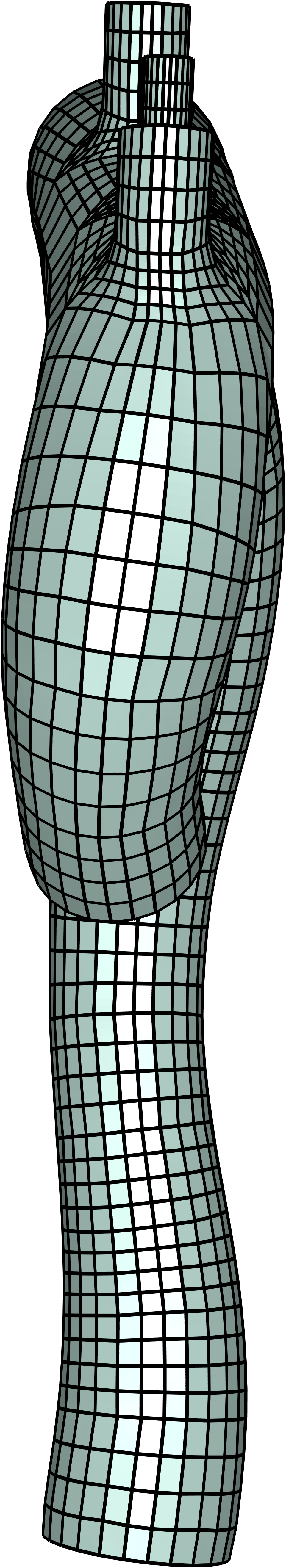}
        \includegraphics[height=3.5cm]{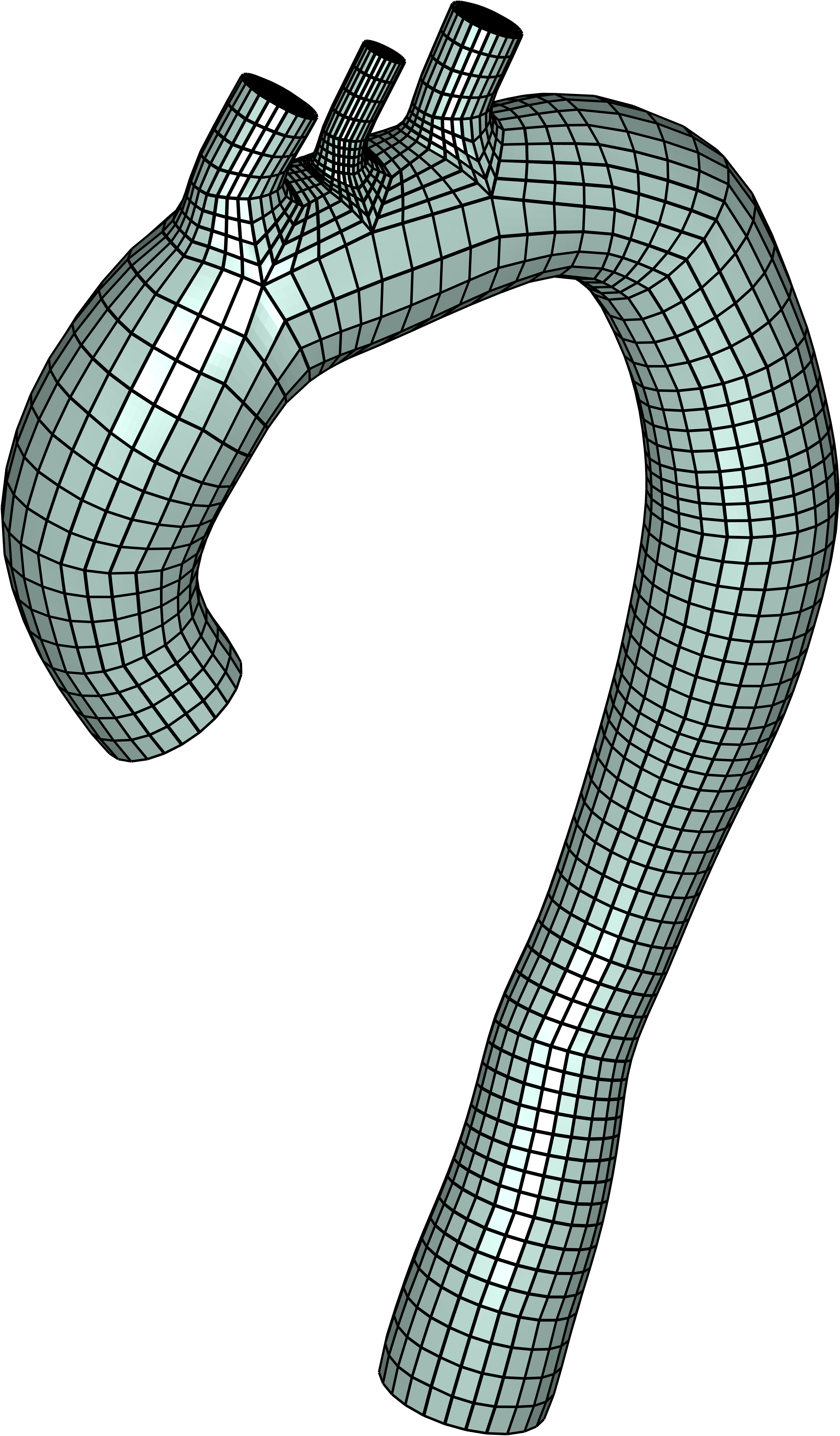}
    }
    \hspace{0.5cm}
    \subfloat[]{
        \includegraphics[height=3.5cm]{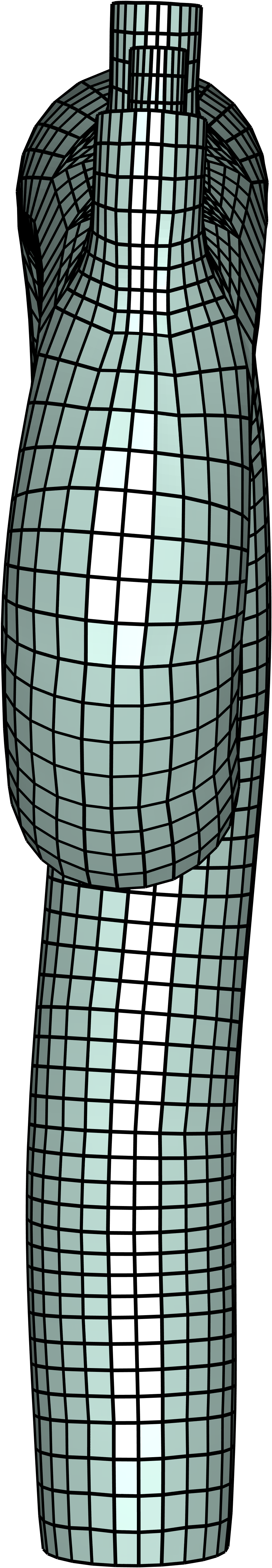}
        \includegraphics[height=3.5cm]{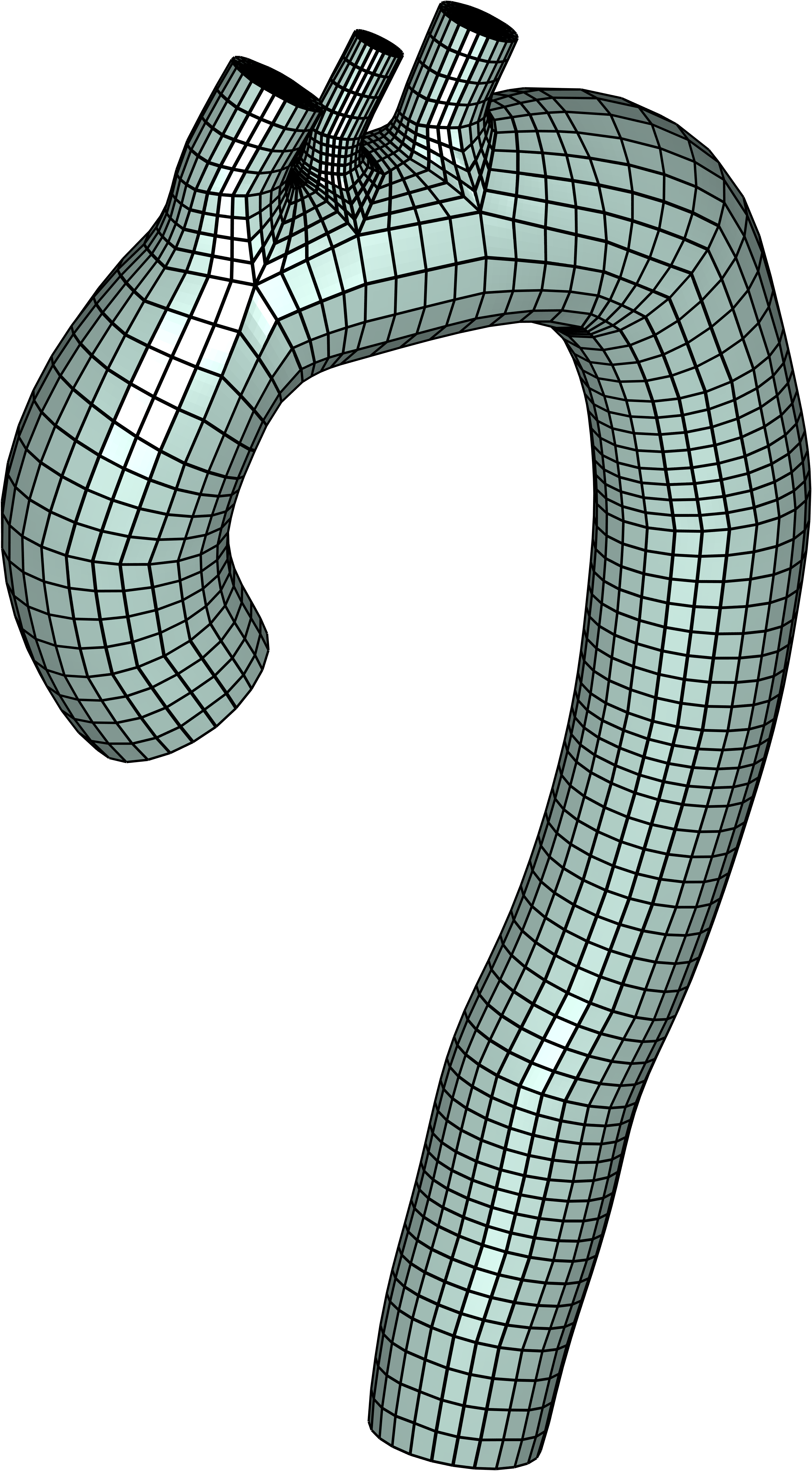}
    }
    \hspace{0.5cm}
    \subfloat[]{
        \includegraphics[height=3.5cm]{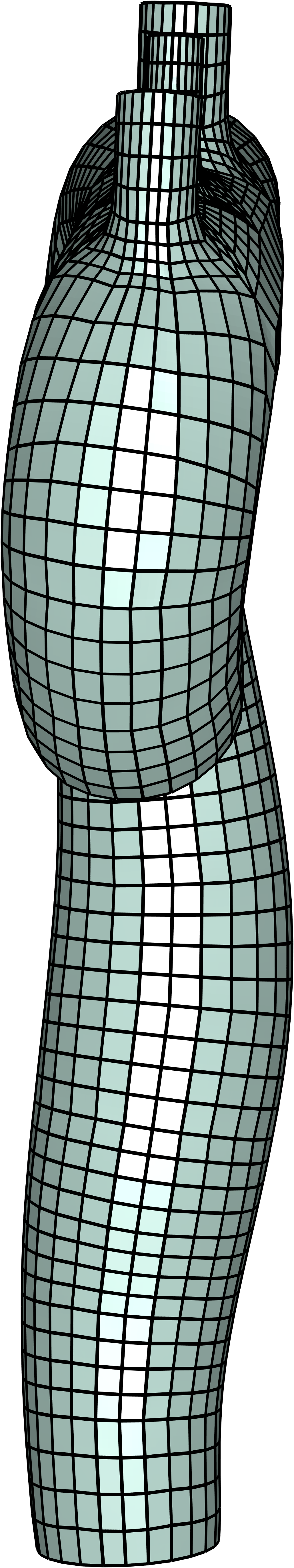}
        \includegraphics[height=3.5cm]{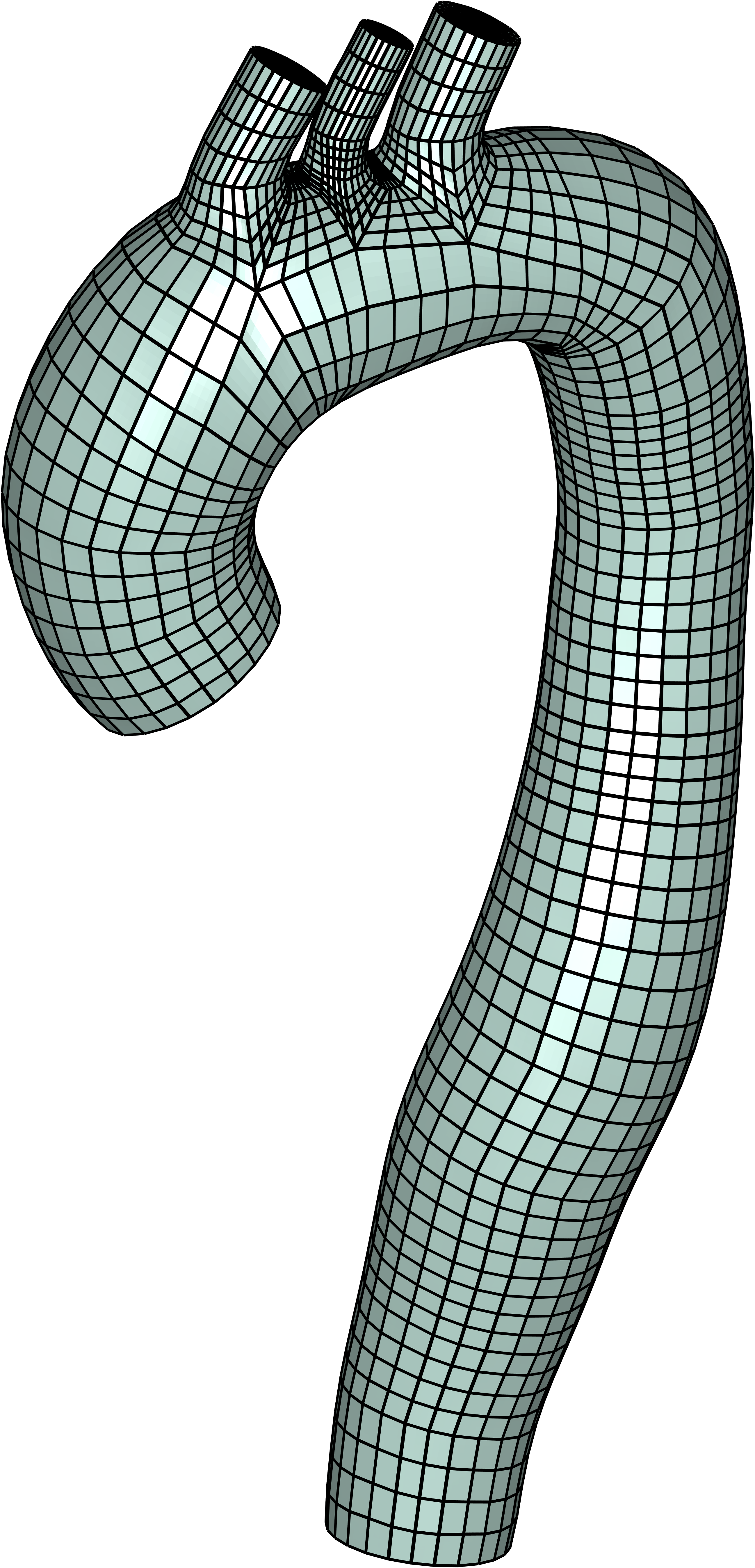}
    }
    \hspace{0.5cm}
    \subfloat[]{
        \includegraphics[height=3.5cm]{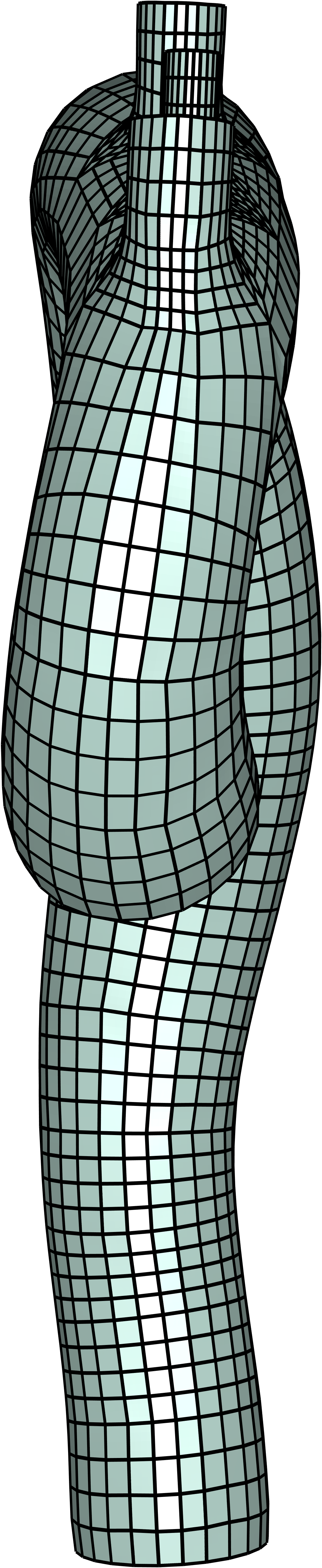}
        \includegraphics[height=3.5cm]{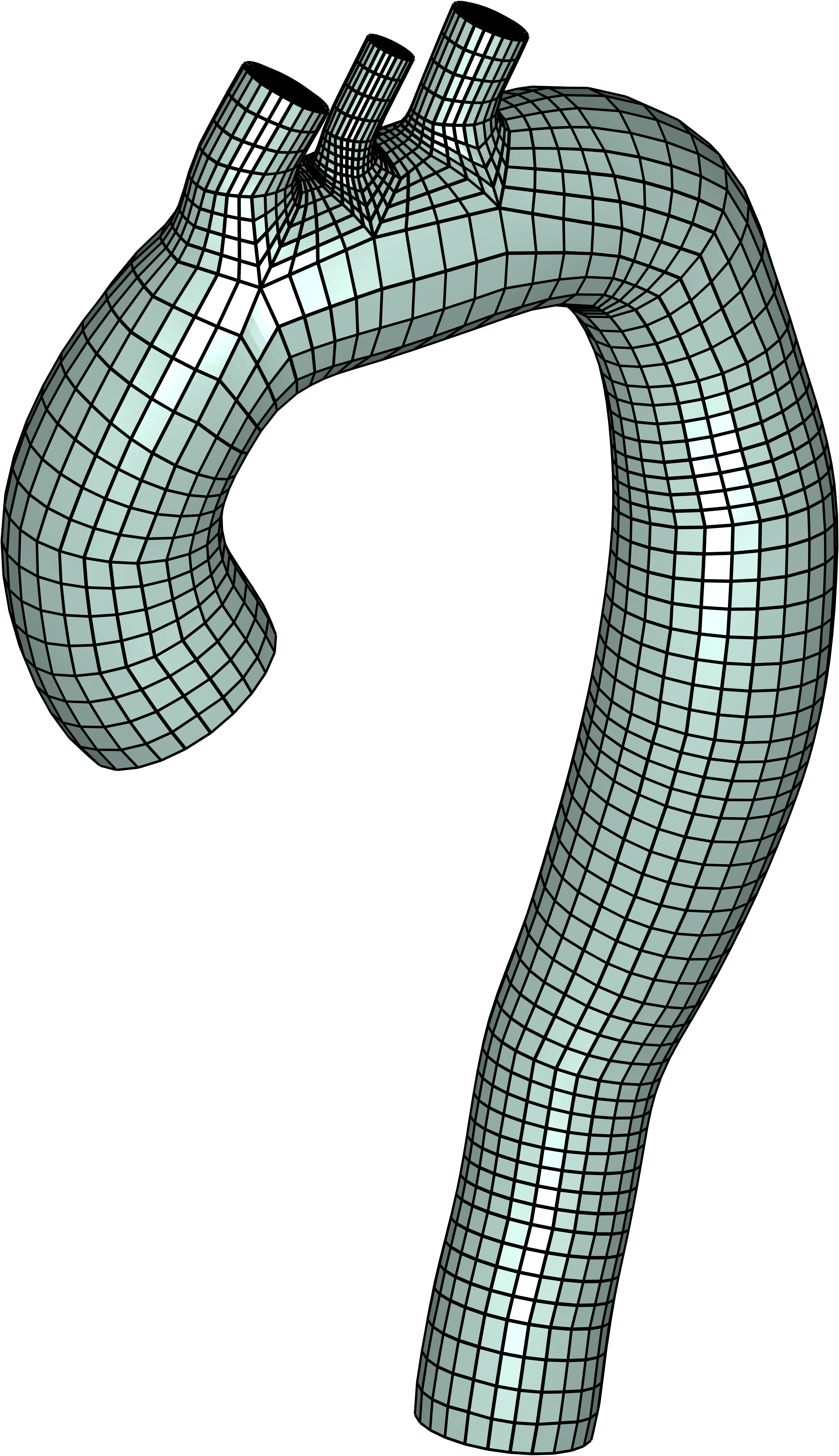}
    }
    \\
    \subfloat[]{
        \includegraphics[height=3.5cm]{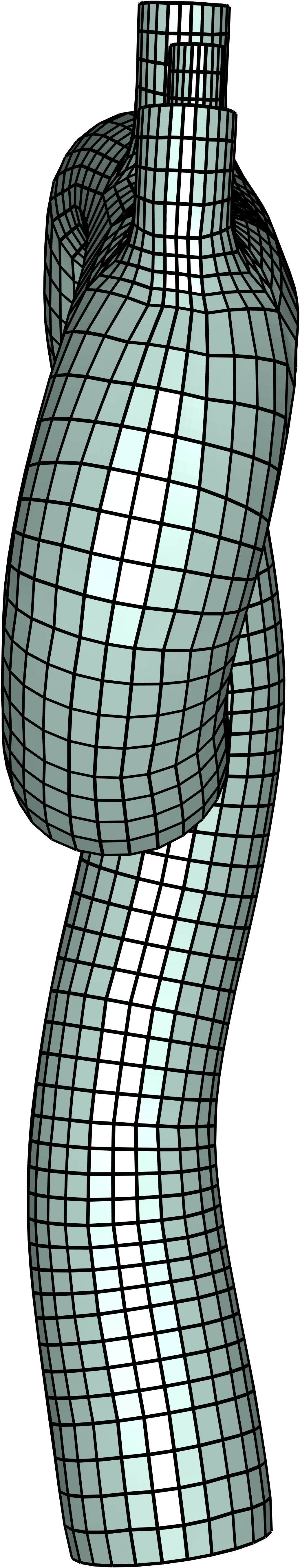}
        \includegraphics[height=3.5cm]{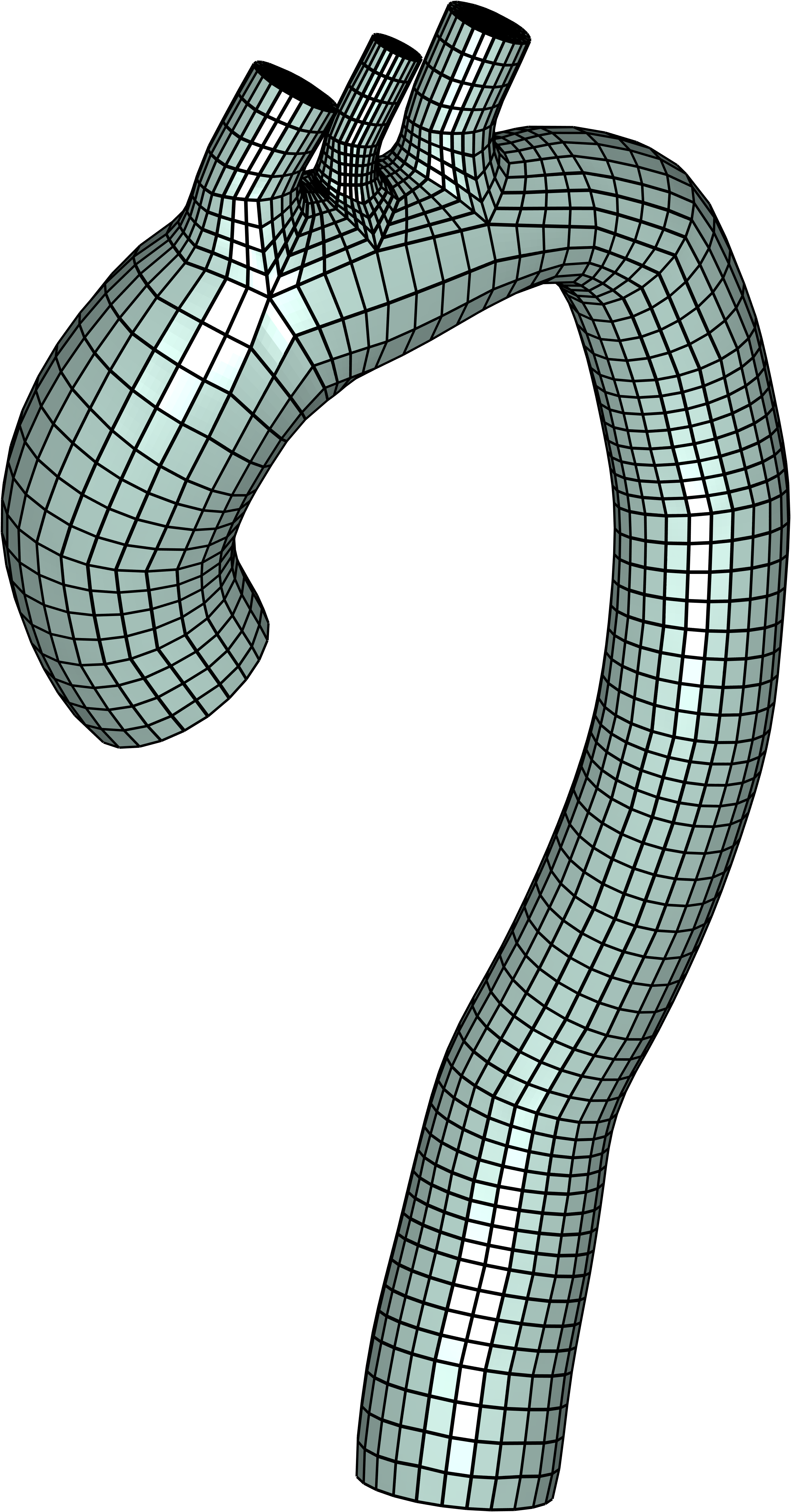}
    }
    \hspace{0.5cm}
    \subfloat[]{
        \includegraphics[height=3.5cm]{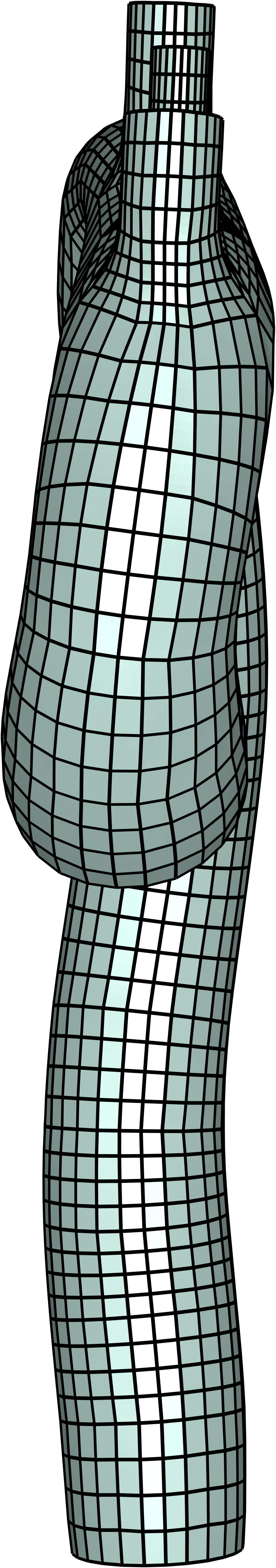}
        \includegraphics[height=3.5cm]{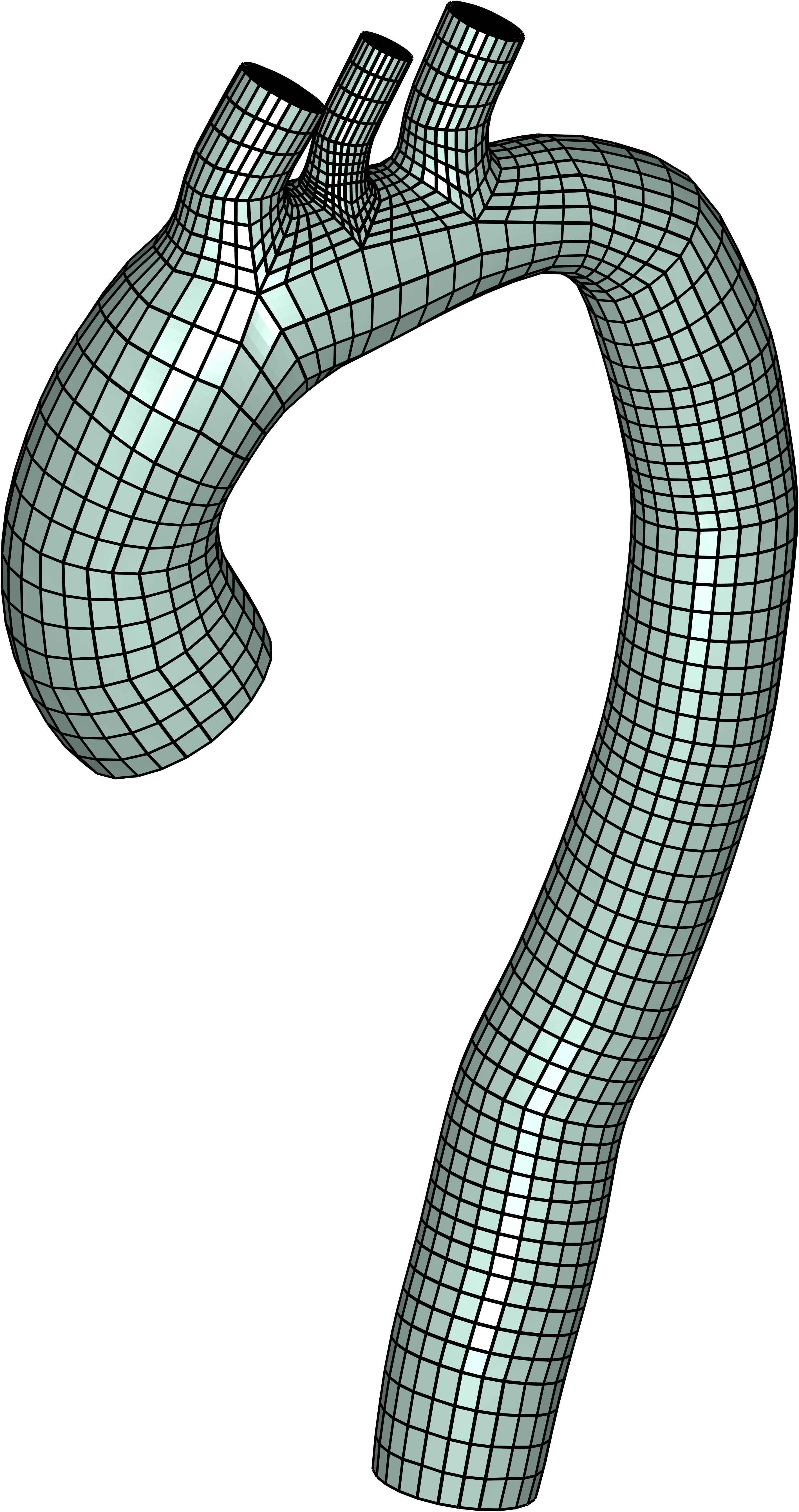}
    }
    \hspace{0.5cm}
    \subfloat[]{
        \includegraphics[height=3.5cm]{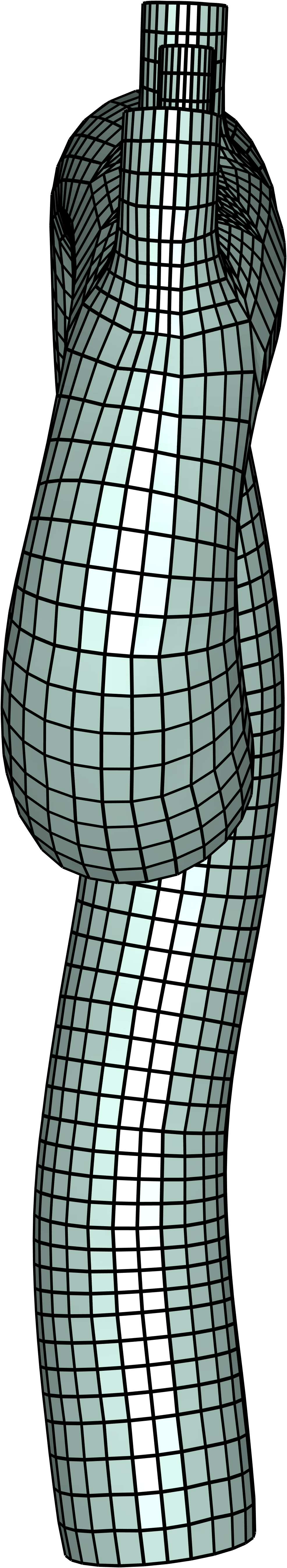}
        \includegraphics[height=3.5cm]{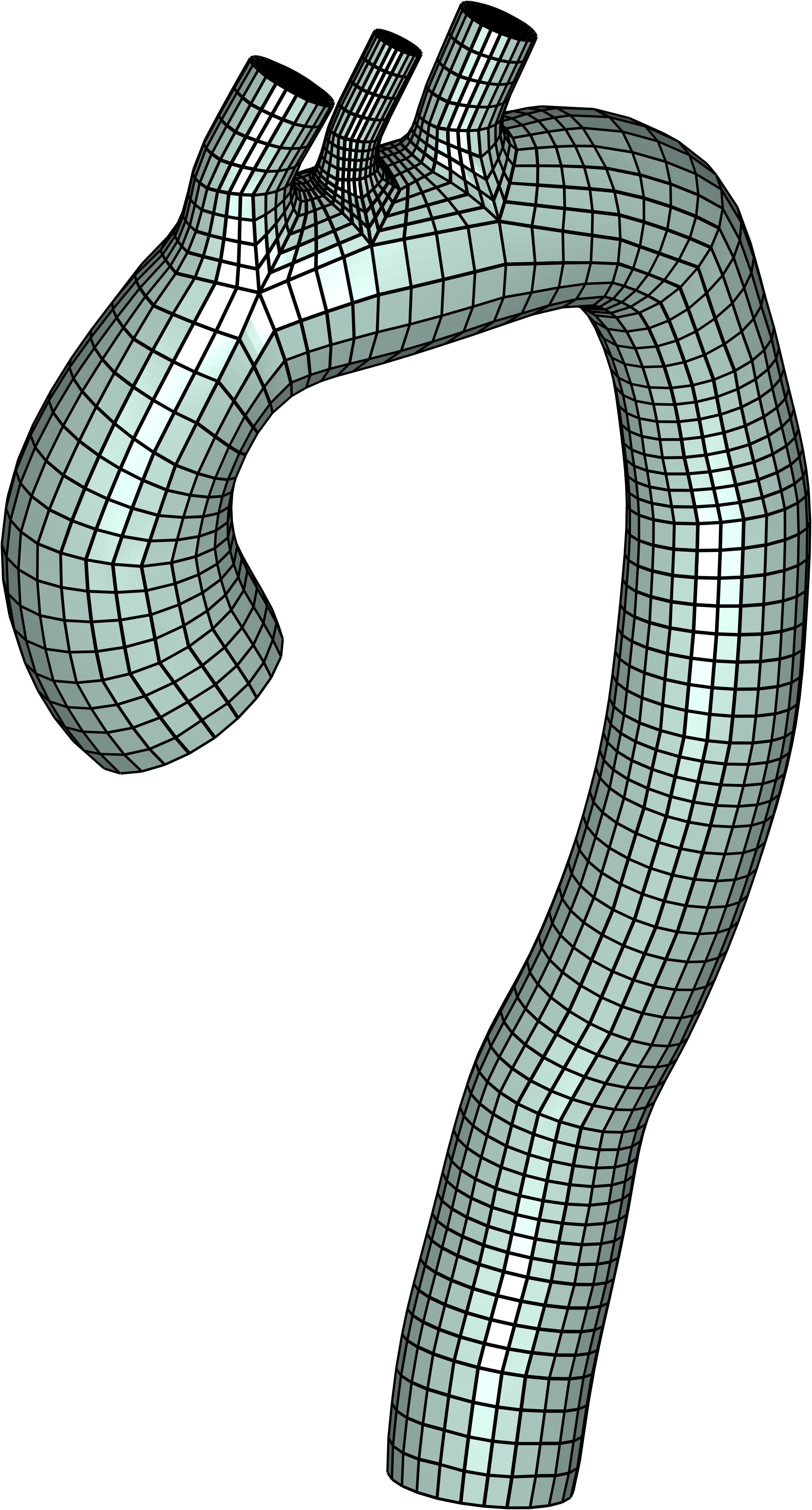}
    }
    \hspace{0.5cm}
    \subfloat[]{
        \includegraphics[height=3.5cm]{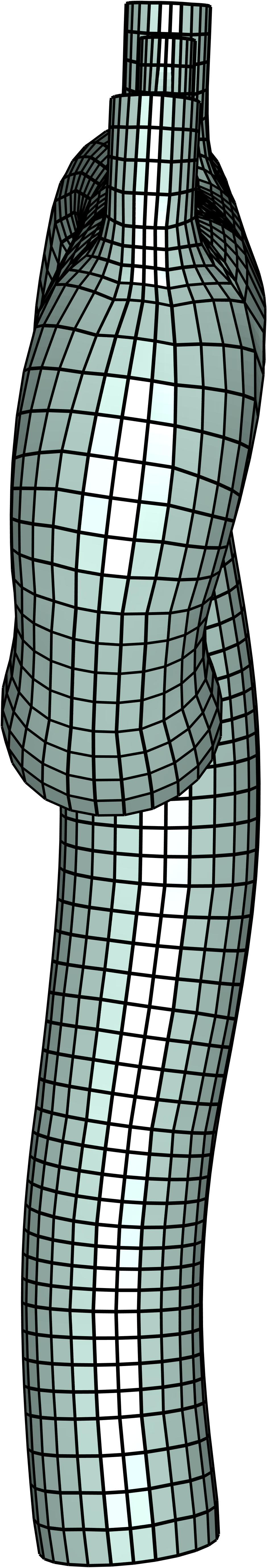}
        \includegraphics[height=3.5cm]{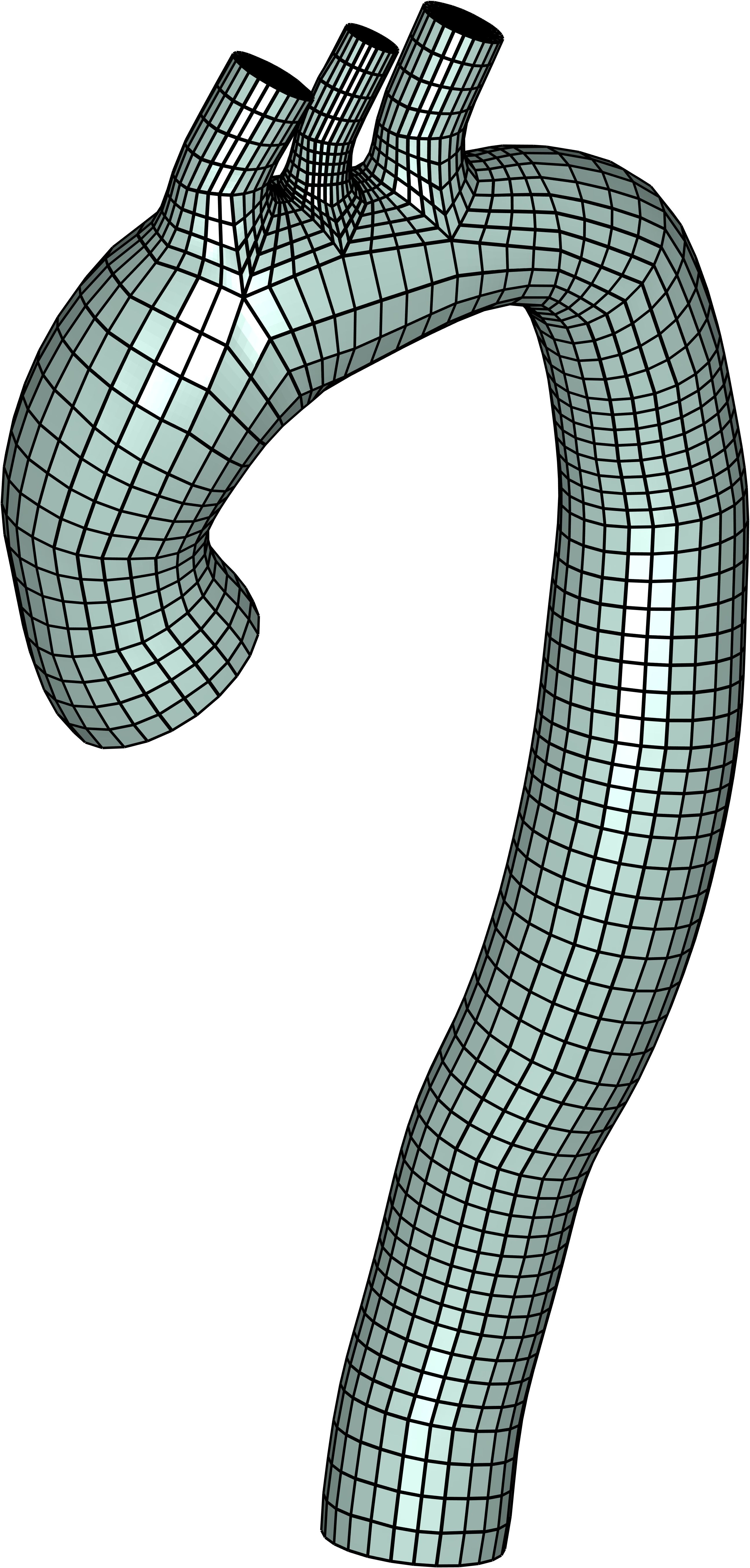}
    }
    \hspace{0.25cm}
    \caption{Examples from the SynthAorta dataset: structured hexahedral meshes of different aortic geometries, in different views. Note that the aortas are scaled for better visibility, but they are of varying heights/lengths as well.}
    \label{fig:DatasetExamples}
\end{figure*}

The entire dataset is recovered from a single patient-specific model~\cite{Bosnjak_2023a}, and a set of parameters shown in Tab~\ref{tab:aorticParametersDistributions}. Several examples from the dataset are illustrated in Fig.~\ref{fig:DatasetExamples}. 

The dataset is stored efficiently since each mesh only needs to keep track of the \(x,\,y \text{ and } z\) values of the nodes, which are stored in a standard \emph{csv} file format. A single connectivity matrix is stored, as it is the same for every mesh. Moreover, we store connectivity matrices for \(3\) coarser mesh levels in the same manner, with \(224, 1\,792,\) and \(14\,336\) elements, respectively. {The main mesh contains \(114 \, 688\) elements.} A simple function to extract the nodes for a coarser mesh into a new file is provided as well. {For efficiency, the data is stored in binary format, using floats for 3D coordinates, and integers for connectivity matrices.}

\subsection{Mesh quality}
In order to verify the suitability of the dataset for simulations, we measure a standard mesh quality metric, the \emph{scaled Jacobian}. A value is obtained for each mesh element, with negative values indicating invalid elements, and values closer to \(1\) the perfect elements. In the examples we produced, the minimum value the median value stays above \(0.84\), and a very small number of elements drops under \(0.5\). This implies simulation-ready mesh quality, a property that is not nearly as easy to robustly achieve in structured hexahedral meshing as it is in unstructured tetrahedral approaches~\cite{Livesu_2016a, Decroocq_2023a}. Naturally, mesh optimization techniques can still be applied if desired.

\subsection{Aortic flow simulations}

We first verify the suitability of the meshes through a convergence study.
Three cardiac cycles $t\in (0,2.34~\mathrm{s}]$ are computed on successively finer grids of the grid with the worst element quality. Relative errors are computed taking the finest resolution as reference, see Tab.~\ref{tab:convergence_results}. We observe convergence of the flow rate and spatial mean pressure (exemplarily given for the outlet at abdominal level), and the WSS in the descending aorta. The WSS is subject to greater errors, since it is a function of the velocity gradient and the unit outward normal. Considering level $l=1$ for all meshes in what follows, we have thus strong indications that the quantities of interest (QoIs) are resolved with engineering precision (flow rates and pressure $<5\%$, WSS $<15\%$). However, similarly strict convergence studies for all cases in the SynthAorta dataset are computationally prohibitive. The resolution required depends on the physics of interest and can hence not be given in general, rendering the multiple grid resolutions provided in the SynthAorta dataset useful also in this sense.

\begin{table}[]
  \centering
  \caption{
  Relative errors in the flow rate $\epsilon_Q$ and spatial mean pressure 
  $\epsilon_{\bar{p}}$ at the outlet at abdominal level, as well as the WSS 
  $\epsilon_\mathrm{WSS}$ in the descending aorta under mesh refinement using mesh levels $l=0,1,2$.
  }
    \begin{tabular}{llll}
    \toprule
    Mesh level & $\epsilon_Q$~[\%] & $\epsilon_{\bar{p}}$~[\%] & $\epsilon_{\mathrm{WSS}}$~[\%] \\
    \midrule
    0             & 20.54           & 3.61 & 99.81       \\
    1             & \phantom{0}1.91 & 0.22 & 11.00       \\
    2 (reference) & \phantom{0}-    & -    & \phantom{0}-\\
    \midrule 
    \end{tabular}%
  \label{tab:convergence_results}%
\end{table}%

\begin{figure*}[t]
    \centering
    \subfloat[]{
        \includegraphics[height=4cm]{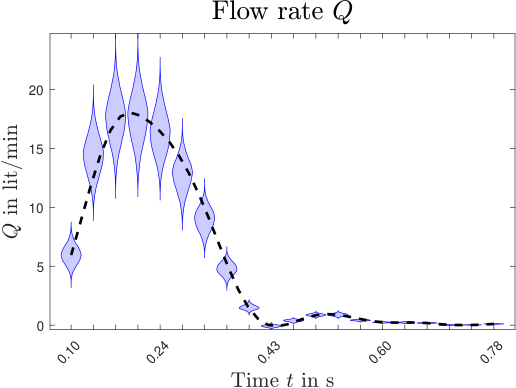}
        \label{fig:violin_flow}
    }
    \hspace{0.0cm}%
    \subfloat[]{
        \includegraphics[height=4cm]{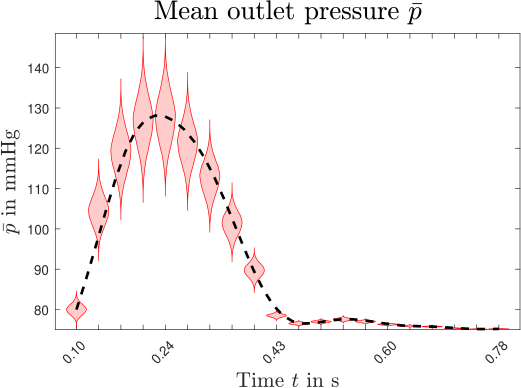}
        \label{fig:violin_pressure}
    }
    \hspace{0.0cm}%
    \subfloat[]{
        \includegraphics[height=4cm]{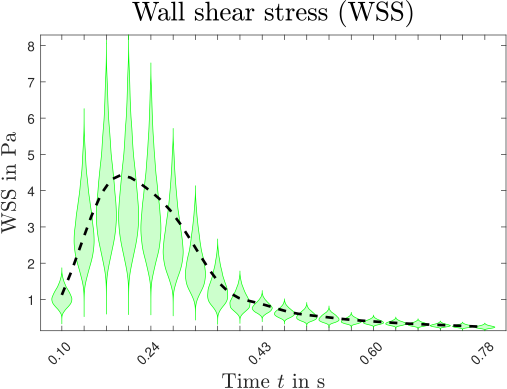}
        \label{fig:violin_wss}
    }
    \caption{Temporal trace of the QoIs in the abdominal aorta: flow rate (left), mean pressure (middle) and spatial average of the wall shear stress (right) in the abdominal aorta. The deviation from the mean over the entire SynthAorta dataset (black dashed line) is largest in the systolic phase, while the pressure and flow show significantly less variation than the WSS.}
    \label{fig:violin_plots}
\end{figure*}

Lastly, we provide hemodynamics QoIs for the entire SynthAorta dataset, where we reduce the interval of interest to the first cardiac cycle $t\in(0,0.78~\mathrm{s}]$, which is admissible given the tuned Windkessel parameters and omitting the first $t\approx0.1~\mathrm{s}$, after which periodicity is already reached (discrepancies lie in the single-digit range comparing QoIs in the time interval $[0.10, \, 0.78]$ with the periodic solution). In order to process massive amounts of fluid flow simulations of adequate resolution, optimized parameters in the simulation tools have to be fine-tuned. Inevitably, some simulations fail to complete if the settings are optimized for performance as necessary in the present case. Such, almost diverging, simulations yield unphysical results, and thus require physical filtering and outlier removal (for a detailed discussion of this dilemma see~\cite{Bosnjak_2025b}). The corresponding results are depicted in Fig.~\ref{fig:violin_plots} showing the QoI distributions in the SynthAorta dataset over the cardiac cycle. We observe that the deviation from the mean over the entire dataset is largest in the systolic phase, while the pressure and flow show significantly less variation than the WSS. This hints at the expected heavy impact of the shape hyperparameters on hemodynamics, while the relative importance of these parameters remain to be identified. Within this work, these results simply serve to demonstrate the applicability of the dataset to real-world physics. It lays the foundation for further investigations centered around sensitivity analysis and uncertainty quantification or model order reduction/surrogate modeling including machine learning techniques. 

\section{Conclusion}
SynthAorta encompasses a comprehensive dataset of synthetic, physiological aorta models, including a centerline, a surface, and a structured hexahedral mesh. The structured meshes make them highly suitable for reduced-order modeling and machine-learning applications.
Our approach is based on using one patient-specific segmented aorta, referred to as the base model, and statistical information from medical literature. We utilize convolution surfaces to create high-quality visual representations and structured meshes for numerical simulations. We have developed a new set of statistical parameters to generate synthetic models of the aorta, ensuring that the parameters closely align with clinical use.

The presented dataset paves the way for realistic analyses of the impacts of the aortic geometry on arbitrary quantities of interest. The parameters are verifiably in physiological intervals, making the comparison of results consistent with other works in the literature, as well as with clinical studies. The convolution surfaces enable proper visualization, whereas the structured hexahedral meshes present a ready-to-use product for performing computational fluid dynamics simulations. A coarse mesh extraction is present as well, enabling up to \(3\) uniform coarsening levels. This is particularly useful for applications where a huge number of simulations needs to be performed, e.g., in uncertainty quantification or sensitivity analysis.

Naturally, the approach comes with certain limitations. We do not claim that such a small parameter set, based on a single aortic model, captures the entire, comprehensive variety of the aortic morphology. More work needs to be invested to make it fully comprehensive. Extensions to certain pathological aortic geometries might be difficult due to the convolution surfaces' inability to render sharp, intricate features. Independent parameter sampling can produce non-physiological geometries, an effect slightly exacerbated by the convolution surface blending, resulting in some examples being discarded. However, the detection of such situations is automated during the convolution surface construction. Finally, clinical validation of the resulting dataset is not in the scope of this paper.

For future work, we aim to extend this functionality to topological domain changes as well, which is rather straightforward for convolution surfaces, as well as analyze the statistical distribution of the generated geometries in the context of common arterial diseases. Specifically, we will explore whether our method can produce geometries resembling coarctations or aneurysms in various aorta segments. When contrasting these synthetic cases with documented instances in the literature, we will evaluate our models' accuracy and clinical significance. This extension will further validate our approach and enhance its applicability in simulating and studying pathological conditions, offering a valuable tool for research and clinical practice.

\section*{Data availability}
The full dataset is available open-source at the repository of the Graz University of Technology~\cite{Bosnjak_2025a}, containing the minimal code to load meshes and centerlines, as well as to export them to the well-known .msh format, enabling visualization and manipulation in the open-source tool Gmsh~\cite{Geuzaine_2009a}. Since the dataset is fixed in the repository, any new developments or additions will be done on the corresponding Github project at https://github.com/domagoj-bosnjak/SynthAorta. Moreover, it contains \(100\) examples from the dataset in the same format.

\section*{Acknowledgements}
The authors gratefully acknowledge the support by Graz University of Technology, Austria for the financial support of the LEAD-project: ``Mechanics, Modelling and Simulation of Aortic Dissection''. R.S. was partially supported by the German research foundation through project no. 524455704 (``High-Performance Simulation Tools for Hemodynamics'') and partially supported by the EuroHPC joint undertaking Centre of Excellence dealii-X, grant agreement no. 101172493.

\bibliographystyle{ieeetr}
\bibliography{BosnjakRefsBibTeX}

\end{document}